# Quasi-solitons in Rydberg atom chains

Aron Kerschbaumer [1], Jean-Yves Desaules [1], Marko Ljubotina [1,2,3] & Maksym Serbyn [1]

Solitons—localized wave packets that travel without spreading—play a central role in understanding transport and properties of nonlinear systems. In quantum many-body systems, however, such robust excitations are typically destroyed by thermalization. Here, we theoretically demonstrate the existence of solitonic excitations in high-energy states of Rydberg atom chains in the regime of strong nearest-neighbor Rydberg blockade. These localized wave packets propagate directionally atop a special class of reviving initial states related to quantum many-body scars and are capable of carrying energy. Exhibiting long coherence times, these states constitute a form of non-ergodic quantum dynamics and can be efficiently implemented on Rydberg atom simulators. In this work, in addition to a phenomenological description of solitons, we identify their counterpart in a classical nonlinear dynamical system, demonstrate their potential use in quantum information transfer, and conjecture their relevance for anomalous energy transport reported in numerical studies of Rydberg atom arrays.

Coherently propagating solitons in nonlinear systems have been realized in numerous classical and quantum platforms, ranging from water waves and optical fibers to Bose-Einstein condensates[1], magnetic materials[2], and superconducting circuits[3]. While many of these realizations occur in continuum media, modern experimental advances enable the study of discrete solitons, realized in particular by arrays of optical waveguides and similar systems[4]. In quantum settings, soliton excitations are known to exist in numerous integrable lattice models[5,6], yet their experimental study is currently limited to dynamics of isolated magnon bound states, which can be regarded as the basic building blocks of magnetic solitons[7].

In our work, we establish the existence of quasi-solitons in the PXP model[8,9], a strongly interacting, chaotic system that corresponds to an approximate description of Rydberg atom arrays[10–12]. They exist in high-energy states of the PXP model and can be efficiently prepared and launched in Rydberg atom simulators. Specifically, we show that quasi-soliton excitations can be created as simple defects on top of homogeneous short-range entangled states that feature coherent revivals[11,12].

These quasi-solitons can carry energy, set by a continuous internal degree of freedom. Multiple quasi-solitons can also be created. This only weakly affects their coherent propagation, except for collisions, which lead to partial decay. Their imperfect nature is in line with the non-exact revivals of the scarred initial states[13,14] they are built upon. This approximate scarred dynamics in a chaotic system makes it difficult to rigorously relate our coherently propagating quasiparticles to solitons in the strict sense. Therefore, we call them quasi-solitons, and establishing a rigorous connection will be a subject of future work. For brevity, we refer to them simply as solitons for the rest of the manuscript; we return to a more detailed discussion of their quasi-solitonic character in the Discussion. These multi-soliton states span an exponentially large manifold of non-thermal, experimentally relevant states. The dynamics of a representative state is shown in Fig. 1 via local observables—the number operator density **a** and energy density **b**—showing uniform soliton propagation on top of the oscillating background. The exponential proliferation of long-lived energy-carrying solitons suggests that they may underlie the unconventional energy transport observed in the PXP model[15] and points to potential applications in quantum-information protocols, which we will demonstrate by launching a Bell pair of oppositely moving solitons, generating long-range entanglement in an experimentally accessible quench. In addition to the phenomenological construction of solitons, we also identify their counterparts in a classical dynamical system, establishing a connection to solitons in classical nonlinear systems.

[1]Institute of Science and Technology Austria (ISTA), Klosterneuburg, Austria. [2]Physics Department, Technical University of Munich, TUM School of Natural Sciences, Garching, Germany. [3]Munich Center for Quantum Science and Technology (MCQST), München, Germany. e-mail: aron.kerschbaumer@ist.ac.at; jean-yves.desaules@ist.ac.at; marko.ljubotina@tum.de; maksym.serbyn@ist.ac.at

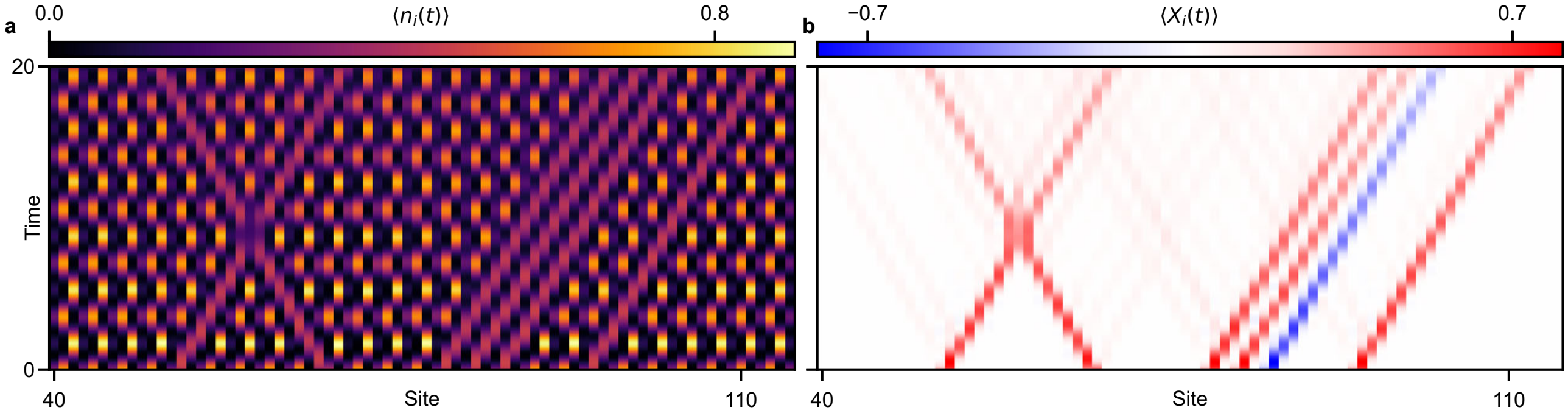

**Fig. 1 | Dynamics of an energy-carrying multi-soliton state.** Solitons are embedded as local defects in the weakly entangled scarred state $|K\rangle$. The solitonic defects are propagating unidirectionally atop the oscillating scar background, as seen in the evolution of **a**, number operator density and **b**, energy density. The collision leads to additional soliton decay. Details of the numerical simulation and the exact form of the states are provided in the Methods section.

## Results

### Model and properties

The dynamics of a Rydberg atom chain can be described as an effective two-level system in which the ground state $|\downarrow\rangle$ is coupled to the excited state $|\uparrow\rangle$ via driving with a homogeneous Rabi frequency $\Omega$ [10–12]. Atoms in excited states $|\uparrow\rangle$ interact with rapidly decaying van der Waals interactions. By tuning an appropriate inter-atom distance, the chain can be brought to the regime of nearest-neighbor Rydberg blockade, characterized by strong nearest-neighbor interactions and negligible interactions beyond. Such an $N$-atom Rydberg chain with nearest-neighbor blockade and periodic boundary conditions can be approximated by the so-called PXP model with the Hamiltonian[9,11]:

$$H = \sum_{i=1}^{N} P_{i-1} X_i P_{i+1}. \tag{1}$$

The projectors onto the ground state $P_i = (|\downarrow\rangle\langle\downarrow|)_i$ enforce a kinetic constraint by preventing neighboring atoms from occupying the excited state simultaneously. The Pauli matrix $X_i = (|\downarrow\rangle\langle\uparrow| + |\uparrow\rangle\langle\downarrow|)_i$ creates Rabi oscillations between the ground and excited state. The model obeys a global chiral symmetry that will play an important role below, generated by an operator $\mathcal{C}$ that upon conjugation changes the sign of the Hamiltonian:

$$\mathcal{C} = \prod_{i=1}^{N} Z_i, \qquad \mathcal{C}H\mathcal{C} = -H, \tag{2}$$

with $Z_i = (|\uparrow\rangle\langle\uparrow| - |\downarrow\rangle\langle\downarrow|)_i$. The PXP model has been intensively studied due to the presence of quantum many-body scars[16–18]—non-ergodic eigenstates embedded in its otherwise thermal spectrum[13,19–22]. In a quantum quench setup, when one follows the unitary evolution generated by Eq. (1) from simple initial states, quantum many-body scars are manifested in approximate periodic revivals for a few, but not generic, initial states. Intuitively, such scarred initial states are located on the periodic trajectory of an effective semiclassical description of the dynamics, and therefore lead to revivals in a quantum quench[23,24]. The most studied scarred initial state is the Néel product state $|\mathbb{Z}_2\rangle = |\downarrow\uparrow\downarrow\uparrow\ldots\rangle$ constructed by repeating the unit cell of two atoms, $|\downarrow\uparrow\rangle$[11,13]. Notably, perturbing the fine-tuned scarred initial states away from the semiclassical time-dependent variational principle (TDVP) trajectory destroys the periodic revivals. Even a single defect inserted into the $|\mathbb{Z}_2\rangle$ state, $|\ldots\downarrow\uparrow\downarrow\downarrow\uparrow\downarrow\ldots\rangle$, disturbs the state within a light-cone that ballistically expands throughout the chain in both directions[25,26]. While this impurity can be confined by a strong staggered potential[25,27] or reflected by a large potential barrier to achieve propagation in a single direction[28], stabilizing it seemingly always requires a strong deformation of the PXP Hamiltonian.

To obtain coherently propagating chiral excitations in the PXP model, we focus on a different scarred 3-site periodic initial state that was obtained in earlier works using semiclassical trajectories[24] and an approximate su(2) algebra[29] (which explains previously reported revivals of the product state $|\mathbb{Z}_3\rangle = |\downarrow\downarrow\uparrow\downarrow\downarrow\uparrow\ldots\rangle$[19] through the large overlap between the two states). This state, with $N$ atoms and $N/3$ unit cells, can be built by repeating the following weakly-entangled 3-site unit cell:

$$|K\rangle = \bigotimes_{k=1}^{N/3} |S\rangle, \qquad |S\rangle = \frac{1}{\tilde{\beta}}|\downarrow\rangle(\beta|\uparrow\downarrow\rangle + |\downarrow\uparrow\rangle), \tag{3}$$

where $\tilde{\beta} = \sqrt{\beta^2 + 1}$ and $\beta = 0.65$.

### Soliton excitations

Coherently propagating defects—which we will refer to as solitons—can be constructed on top of the 3-site periodic initial state $|K\rangle$. The local Hilbert space of allowed 3-site cells that obey the constraint, when surrounded by $|S\rangle$ cells, has dimension 3, and can be spanned by the scarred cell $|S\rangle$, along with the left- and right-moving defect cells $|L\rangle$ and $|R\rangle$. These 3-site defect cells can be determined by requiring (i) the first site to be in the ground state to avoid blockade violation, (ii) $\{|S\rangle, |L\rangle, |R\rangle\}$ being orthonormal to each other and (iii) to satisfy the conjugation condition under chiral symmetry, $|L\rangle = \mathcal{C}|R\rangle$.

Among the one-parameter family of solutions satisfying the above conditions, we find that the following leads to the most pronounced coherent directional propagation, see Fig. 2a:

$$|L,R\rangle \;=\; \frac{1}{\sqrt{2}\tilde{\beta}}|\downarrow\rangle(\mp\tilde{\beta}|\downarrow\downarrow\rangle + i|\uparrow\downarrow\rangle - i\beta|\downarrow\uparrow\rangle), \tag{4}$$

with $\beta, \tilde{\beta}$ being the numerical constants defined after Eq. (3). Below, we will consider the dynamics of initial states formed by $|S\rangle$ unit cells with the insertion of $|L, R\rangle$ defects. For brevity, we denote such states as $|K^{\mathbf{l},\mathbf{r}}\rangle$, where the disjoint sets **l** and **r** specify locations of cells with $|L\rangle$ and $|R\rangle$ states,

$$|K^{\mathbf{l},\mathbf{r}}\rangle = \bigotimes_{j=1}^{N/3} \begin{cases} |L\rangle_j & \text{for } j \in \mathbf{l}, \\ |R\rangle_j & \text{for } j \in \mathbf{r}, \\ |S\rangle_j & \text{otherwise}. \end{cases} \tag{5}$$

A single unit cell $|R\rangle$ initialized in the background of the scarred initial state $|K\rangle$ shows coherent propagation to the right under the unitary evolution with the PXP Hamiltonian, as is visualized in Fig. 2b. This propagation is visible in the expectation value of the Rydberg atom number operator, defined as $n = |\uparrow\rangle\langle\uparrow|$, shown for the time evolved state $\left|K^{\varnothing,\{j\}}(t)\right\rangle = e^{-iHt}\left|S\ldots SSR_jSS\ldots S\right\rangle$ with a single $|R\rangle$ state inserted near the middle. Far away from the initial location of the $|R\rangle$ cell, the dynamics follow periodic revivals. Therefore, in Fig. 2c we show the difference between the expectation value of the number operator for the time-evolved states with and without the right-moving soliton,

$$\langle n_i(t)\rangle_\Delta = \langle K^{\mathbf{l},\mathbf{r}}(t)|n_i|K^{\mathbf{l},\mathbf{r}}(t)\rangle - \langle K(t)|n_i|K(t)\rangle. \quad (6)$$

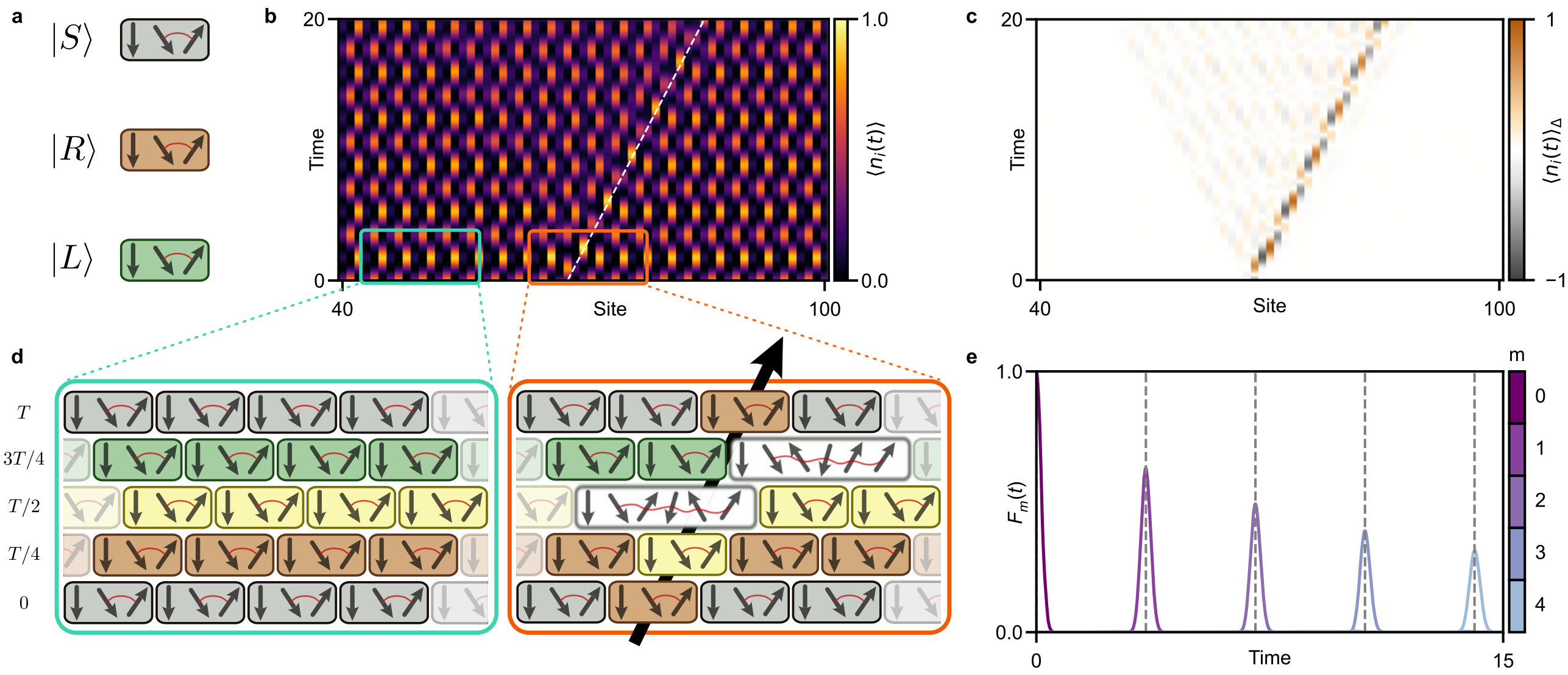

**Fig. 2 | Dynamics of a single right-moving soliton initialized in the scarred state $|K\rangle$ at sites 67 − 69 in a 150-site Rydberg atom chain. a** Legend for scarred cell $|S\rangle$ and soliton cells $|R\rangle$ and $|L\rangle$. The first site in each cell is unexcited, while the second and third sites form an entangled pair. This entangled pair differs for each cell, as indicated by the different colors of the cells, and its exact form is listed in Eqs. (3) and (4). **b** The soliton appears as a mobile defect that locally disrupts the homogeneous precession of the number operator; its trajectory is highlighted by the dashed white line. **c** The difference in the number-operator expectation values between the states with and without the soliton indicates a density perturbation that predominantly moves to the right. **d** (Left) Schematic dynamics of scarred unit cells, $|S\rangle$, shown by gray color within the revival period $T$. (Right) The phase-shifted unit cell from the quarter-periods of the scarred dynamics (orange color) corresponding to $|R\rangle$ in Eq. (4) is embedded within the $|S\rangle$ background. It coherently moves one cell to the right over each revival period $T$, corresponding to a right-moving soliton. Green unit cells at $3T/4$ correspond to left-moving soliton $|L\rangle$, and the state of remaining yellow and white cells is discussed in Supplementary Note 1[30]. **e** Peaks in the translation fidelity (7) for the subsystem $A = \{61, 62, \ldots, 90\}$ at multiples of $T$ indicate near-coherent rightward translation of the entire wave function. Details of numerical simulation and fidelity extraction are available in the Methods section.

Figure 2c confirms that the $|R\rangle$ defect gives a right-moving soliton, as its presence does not lead to the typical symmetric light-cone spreading both to the left and right, but instead affects the state of the system to the right.

The phenomenological explanation of the mechanism by which the $|R\rangle$ impurity forms a right-moving soliton is displayed in Fig. 2d. There, we first sketch the dynamics of the scarred initial state $|K(t)\rangle$ over one period $T \approx 3.52$. Notably, the squared absolute value of the overlap of the state $|R\rangle$ identified in Eq. (4) with the unit cells of the $|K(T/4)\rangle$ state is 0.98 (the same holds for the overlap of $|L\rangle$ and the cell at $3T/4$). Therefore, the creation of the right-moving soliton corresponds to a phase shift of a particular unit cell in the $|K\rangle$ initial state by a quarter period, as shown in the right panel of Fig. 2d. This fast-forwarded unit cell naturally gets shifted to the right after time $T/4$. After this, the state of the soliton can no longer be viewed as a 3-site cell, as it forms an entangled state with the adjacent cell to the left. See Supplementary Fig. 1[30] for additional details. By the full period, the soliton cell $|R\rangle$ reemerges, but now shifted by 3 sites to the right. A similar picture holds for the insertion of the $|L\rangle$ cell into the $|K\rangle$ state, which, however, propagates to the left, corresponding to the left-moving soliton. The left propagation of $|L\rangle$ under unitary dynamics follows from the chiral symmetry of the Hamiltonian (2) and the relation $|L\rangle = \mathcal{C}|R\rangle$ (see Supplementary Note 1[30]).

To quantify how coherently the solitons move between different unit cells under unitary dynamics, we define a family of measures called translation fidelities, labeled by index $m$ specifying the number of unit cells each right-/left-moving soliton has shifted towards the right/left:

$$F_m(t) = \left|\left\langle K^{\mathbf{l}-m,\mathbf{r}+m}\middle|K^{\mathbf{l},\mathbf{r}}(t)\right\rangle\right|^2, \quad (7)$$

with $\mathbf{r} + m := \{j + m | j \in \mathbf{r}\}$. As shown in Fig. 2e, the translation fidelities for the initial state with a single right-moving soliton show large peaks occurring at $t = mT$ for $F_m(t)$. As a result of the slowly scrambling scarred background and the slow decay of the solitons, the fidelities $F_m(t)$ decrease with increasing $m$. Nevertheless, for $m = 4$ at $t \approx 14$ in a large 30-site system, the fidelity remains at 30%, indicating coherent transport of $|R\rangle$ over 12 sites. Even as fidelity diminishes locally, the soliton still shows coherent propagation, as seen in Fig. 2c. This is in stark contrast to the rapid thermalization of generic states (see Supplementary Fig. 2[30]). Long-lived coherence and the preservation of the initial information were previously known for scarred dynamics. However, the soliton state is spatially inhomogeneous, orthogonal to the scarred state $|K\rangle$, and undergoes spatial translation, therefore representing a distinct class of non-ergodic behavior in highly excited states.

## Soliton stability and collisions

After establishing individual right- and left-moving solitons, we study the dynamics of initial states with multiple solitons in Fig. 3. To this end, we first consider states with $k$ right-moving solitons $|R\rangle$, each consecutive pair of solitons separated by two cells $|S\rangle$, embedded in our scarred state $|K\rangle$ in the form of $|\ldots SRS\ldots SRS\ldots\rangle$. Figure 3a shows the value of the peaks in the translation fidelities (7) at the corresponding multiple of the reviving period $T$. The data labeled with $k = 0$ corresponds to the dynamics of the scarred initial state $|K(t)\rangle$, whose revivals feature weak but nevertheless exponential decay with time. Introducing isolated solitons into the initial state $|K\rangle$ enhances this decay with time, by an amount approximately proportional to the number of solitons. Note that, although the smallest peak in fidelity revivals is only about 1% for the four-soliton state, this is still a very large fidelity for a region of 60 sites. Accordingly, the propagation of solitons remains clearly visible in the expectation value of the number operator, see Fig. 3c.

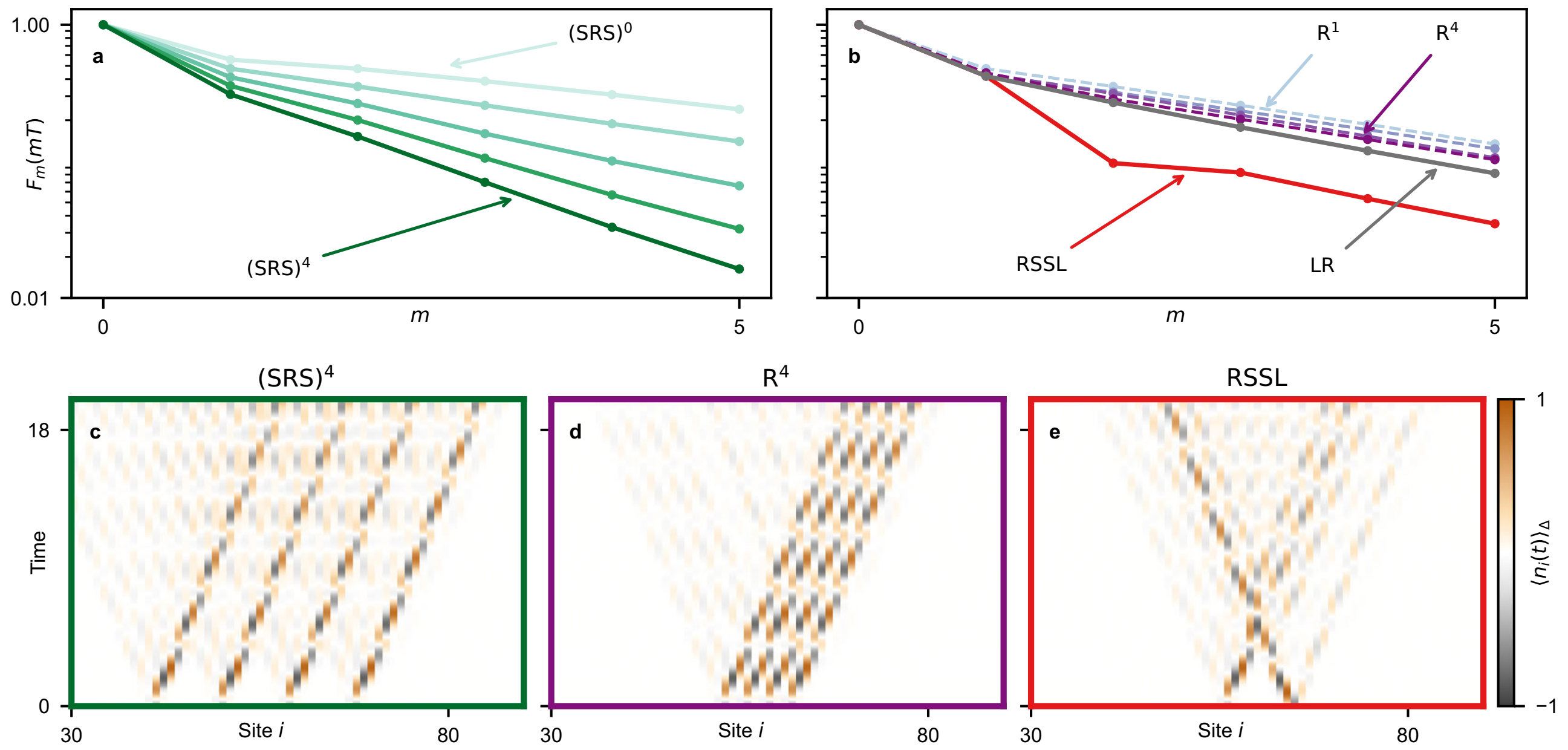

**Fig. 3 | Decay of initial multi-soliton states under unitary dynamics. a** For isolated right-moving solitons, the decay rate is proportional to the number of solitons. **b** In contrast, for initial states where right-moving solitons are placed side-by-side, the decay rate is nearly independent of the number of solitons, implying that domain walls between $|S\rangle$ and soliton cells are responsible for the enhanced decay. **c**, **d** The expectation value of $\langle n\rangle_\Delta$, Eq. (6), also confirms that the initial four-soliton state with isolated solitons features less coherent propagation compared to the dynamics of the four-soliton bulk excitation. **e** The collision of right-/left-moving solitons in the time interval $t \in [T, 2T]$ leads to a rapid decrease in the translation fidelity visible in **b**. Fidelities are shown for the subsystem $A = \{31, 32, \ldots, 90\}$, see Methods for details.

In contrast to the isolated multi-soliton state, for the initial state where right-moving solitons are adjacent to each other in the form of $|RR\rangle$, $|RRR\rangle$ and $|RRRR\rangle$, the fidelity suppression is almost independent of the number of solitons, see Fig. 3b. The coherent propagation of solitons is also clearer in the number operator dynamics in Fig. 3d. The contrasting behavior in fidelity of multi-soliton states in Fig. 3a-b suggests that additional decoherence originates not from the solitons, but from the domain walls separating the $|R, L\rangle$ soliton and scar cells $|S\rangle$. This is consistent with the identification of soliton cells with phase-shifted unit cells of the scarred initial state: phase-shifting the entire scarred state by $T/4$ should give no additional decay, which is equivalent to the $|RR\ldots\rangle$ state.

In Fig. 3b, we also show the additional decay from the collision between solitons propagating in opposite directions. When two counter-propagating solitons are created in a non-colliding configuration, $|\ldots LR\ldots\rangle$, translation fidelities show smooth exponential decay with time. In contrast, the colliding configuration $|\ldots RSSL\ldots\rangle$ displays a sudden drop in fidelity (see Fig. 3b), which is also visible in the dynamics of the number operator in Fig. 3e.

Finally, we quantify the lifetime of a single soliton. After the initial sharp drop, the translation fidelities in Fig. 3a show a consistent exponential decay for $t \gtrsim T$. Therefore, we extract the exponential decay rate between the stroboscopic peaks $T$ and $5T$. To isolate the local quasi-soliton decay from the extensive decay of the scarred background, we calculate the translation fidelity ratio $R(m) = F^{(1)}_m(mT)/F^{(0)}_m(mT)$, where $F^{(1)}_m$ corresponds to the single quasi-soliton state $(SRS)^1$ and $F^{(0)}_m$ corresponds to the scarred state $(SRS)^0$. This quantity is insensitive to system size and captures the local soliton-induced decay. Assuming $R(m) \propto e^{-\gamma_{\rm sol} mT}$ in this window, we estimate the decay rate as $\gamma_{\rm sol} = -\ln[R(5)/R(1)]/(4T)$ and define the soliton lifetime as $\tau_{\rm sol} = 1/\gamma_{\rm sol}$. Using the values of our numerical simulations shown in Fig. 3a, we get a lifetime of $\tau_{\rm sol} \approx 36$.

## Energy transport via solitons

The discussed left- and right-moving solitons $|L, R\rangle$ have—as the scarred initial state—zero energy density, $\langle K^{\mathbf{l},\mathbf{r}}|P_{i-1}X_iP_{i+1}|K^{\mathbf{l},\mathbf{r}}\rangle = 0$ for any $i$ and $\mathbf{l}$, $\mathbf{r}$ (see Supplementary Note 2[30]). Next, we generalize $|L, R\rangle$ solitons into a continuous family of energy-carrying solitons $|L^\alpha, R^\alpha\rangle$, by forming a superposition of the original $|L, R\rangle$ cell and the scar unit cell $|S\rangle$ labeled by an angle $\alpha$,

$$|R^\alpha\rangle = \cos\alpha|R\rangle + \sin\alpha|S\rangle, \quad \alpha \in [-\frac{\pi}{2}, \frac{\pi}{2}). \tag{8}$$

Such unit cells carry a total energy of $\epsilon_\alpha = \sin(2\alpha)(1+\beta)/\sqrt{2(1+\beta^2)}$ (see Supplementary Note 2[30] for details). The left-moving state $|L^\alpha\rangle$ is defined analogously, and carries energy $-\epsilon_\alpha$ due to chiral symmetry. In what follows, we focus on the dynamics of solitons with the maximal (minimal) energy value, denoted as $|R^\pm\rangle = |R^{\pm\pi/4}\rangle$ (see Fig. 4a) and $|L^\pm\rangle = |L^{\mp\pi/4}\rangle$. Since the energy soliton cell $|R^\pm\rangle$ is a linear superposition of $|S\rangle$ and $|R\rangle$ states, we expect that $|R^\pm\rangle$ propagates identically to $|R\rangle$.

The coherent propagation of energy-carrying solitons is shown in Fig. 1 via the expectation values of the number operator density **a** and the local energy density **b**, denoted as $\langle X_i(t)\rangle = \langle\psi(t)|P_{i-1}X_iP_{i+1}|\psi(t)\rangle$. The solitons carry energy over long distances with a weak residual decay. The collision between $|R^+\rangle$ and $|L^+\rangle$ leads to an additional suppression in their energy density, but no detectable phase shift occurs[6]. In Supplementary Note 2[30], we present more detailed data with respect to the slow decay of multi-soliton states, and furthermore, we show how solitons can be used to construct different homogeneous scarred initial states with tunable energy density.

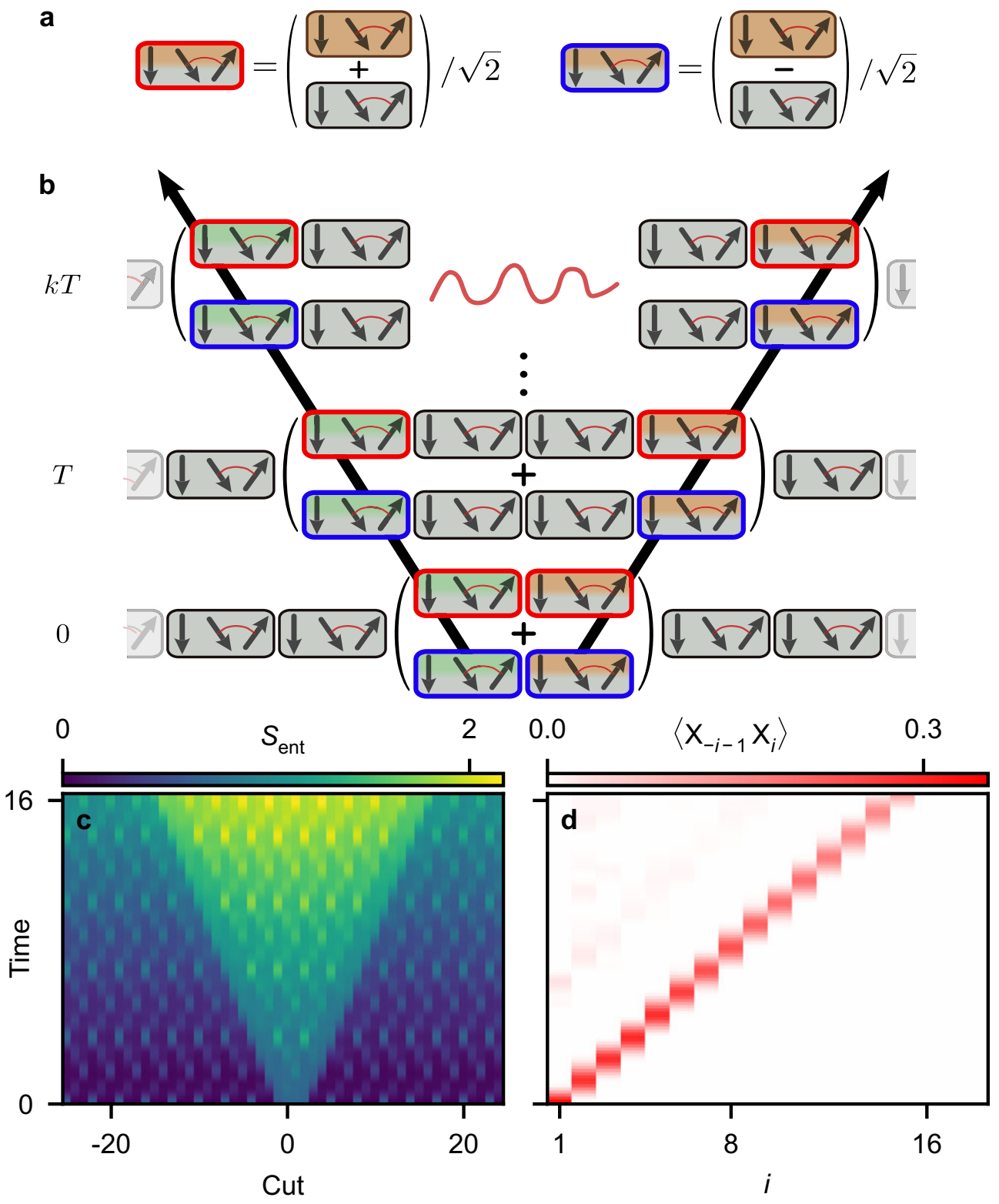

**Fig. 4 | Time evolution of an entangled Bell pair of left- and right-moving solitons embedded in the $|K\rangle$ state at sites {49, …54} in a chain with 102 sites.** **a** Right-moving energy-soliton construction of $|R^+\rangle$ (red) and $|R^-\rangle$ (blue). Left-moving cells $|L^\pm\rangle$ are constructed identically by replacing the brown $|R\rangle$ cells by the green $|L\rangle$ cells. **b** Schematic illustration of the dynamics at integer multiples of the revival period $T$. **c** Bipartite entanglement entropy $S_{ent}$ for different cuts across the chain (with zero corresponding to the link between the original Bell pair cells) shows the spreading of an entanglement cone as the solitons move apart. **d** The connected energy density correlation function, $\langle X_{-i-1} X_i \rangle_c$, exhibits positive correlations along a trajectory with a slope matching the inverse soliton velocity, thus providing an experimental way of confirming long-range quantum information transfer via solitons.

## Information transmission

The coherent propagation of solitons with an internal degree of freedom that corresponds to the energy offers a way to transport information through the Rydberg chain and access it at times that are multiples of the revival period $T$. To this end, we use the orthogonal states $|R^\pm\rangle$ and $|L^\pm\rangle$ to create a Bell pair of left and right-moving solitons, $(|L^+R^+\rangle + |L^-R^-\rangle)/\sqrt{2}$ embedded in $|K\rangle$, see Fig. 4b, traveling in opposite directions without collision. Figure 4c shows the evolution of the bipartite entanglement entropy $S_{ent} = -\mathrm{Tr}[\rho_A \ln \rho_A]$ for different cuts through the chain. The clearly visible light cone spreads in Fig. 4c from the cut position at 0, chosen to be in-between the original soliton Bell pair. To highlight the feasibility of experimental observations, Fig. 4d shows the dependence of the connected energy density correlation function between sites that are mirror-symmetric with respect to the cut position at 1, $\langle X_{-i-1}X_i\rangle_c = \langle X_{-i-1}X_i\rangle - \langle X_{-i-1}\rangle\langle X_i\rangle$. Since the soliton cells each move with a constant velocity $v_{sol}$, the maximum of the correlation function $X_{-i-1}X_i$ occurs at stroboscopic times $t = k/v_{sol}$ at $i = 1 + k$ with $k \in \mathbb{N}$, forming a straight line in Fig. 4d. This soliton-based quantum information transmission procedure is readily implementable in current-generation Rydberg atom setups without requiring site-dependent couplings as in other proposed methods[31]. We note that our demonstrated information transmission is meant only as a proof of principle. Our key point is that quantum information can be transported coherently by quasi-solitonic excitations even though the system is chaotic. By contrast, ref. 32 develops a model for universal quantum computation in Rydberg arrays using global driving. In their model, quantum information is also transported along the chain, but via a pulse-cycle protocol rather than continuous time evolution under a time-independent Hamiltonian.

## Variational description

After demonstrating the existence of solitonic excitations in exact unitary dynamics, we turn to their variational description using an MPS manifold (see Eq. (11)), which intuitively corresponds to the projection of a spin-1/2 coherent state onto the constrained subspace of the PXP model. Projection of exact unitary dynamics onto such a variational manifold with TDVP results in a first-order system of nonlinear differential equations with a chain of $N$ sites leading to $2N$ classical angles $\theta_i$, $\phi_i$, describing the $i$-th spin rotation on the Bloch sphere.

Notably, such an approach was earlier considered only with small unit cells (up to four sites[23,24,29]). We generalize the TDVP equations of motion to an infinite chain with a generic unit cell of $N$ sites (see Supplementary Note 4[30]; see also an independent recent work[33]). When constrained to the subspace with zero energy expectation value, $\phi_i = \pi/2$, our equations of motion take a simple form

$$\partial_t \theta_1 = -\cos\theta_2 - \sin^3\theta_1 \cot\theta_N \frac{\langle n_N\rangle}{\langle n_1\rangle}, \tag{9}$$

where $\langle n_i\rangle$ is the expectation value of the number operator on site $i$ of the unit cell, see Eq. (15) in Methods. The value of $\partial_t\theta_j$ for the remaining angles can be obtained by applying cyclic permutations, 1 → 2 → 3…. Equations (9) are nonlinear and non-local, since the evolution of a given angle explicitly depends on all other angles via the expectation values of the number operators.

This variational description in terms of angles $\theta_i$ captures the homogeneous scarred state $|K\rangle$[24] and the states $|K^{\mathbf{l},\mathbf{r}}\rangle$. At the same time, dynamics generated by Eq. (9) from the initial state with zero-energy solitons $|K^{\mathbf{l},\mathbf{r}}\rangle$ lead to their rapid decay, while the energy solitons show better stability (see Supplementary Fig. 9[30]). The discrepancy between TDVP and exact unitary dynamics is not surprising, as the projection makes TDVP equations nonlinear and may lead to disagreement with unitary evolution. Therefore, we search for solitons directly in the TDVP equations of motion. Numerical optimization of a quantity similar to translation fidelity by three sites (see Supplementary Note 4[30]) results in angles $\theta^S_{1,2,3}$ and $\theta^R_{1,2,3}$ listed in "Methods". The homogeneous state described by the 3-site unit cell with angles $\theta^S_{1,2,3}$ is very close to a point on the trajectory of $|K\rangle$. The defect cell deviates more strongly from our previously reported right-moving soliton; however, Fig. 5a reveals a dynamics of $\langle n_i(t)\rangle_\Delta$ that is very similar to the one observed in exact unitary dynamics (see Supplementary Note 4[30] for more direct comparison). Moreover, when tracking dynamics of individual classical variables $\theta_i$, shown in Fig. 5b,c, we observe that they display periodic motion along approximately closed trajectories. The passing of a soliton through a given site disturbs this motion briefly, after which the angle returns to a replica of the same trajectory, see Fig. 5c. Note that the simple behavior of angles that are not physical observables translates into more complex behavior of local expectation values, such as $\langle n_i\rangle$. Finally, we note that the capability of TDVP to describe the dynamics faithfully depends on the leakage of the trajectory out of the given variational manifold. While we do not explicitly compute the leakage, previous work has shown that the leakage of the discussed scarred trajectory (without a soliton) is small[24], suggesting that TDVP can describe its dynamics. Moreover, in Supplementary Fig. 4[30] we show that the entanglement growth of the purely scarred state and of the state with a single solitonic defect is comparably slow, which supports the expectation that the same variational manifold provides a good description in both cases.

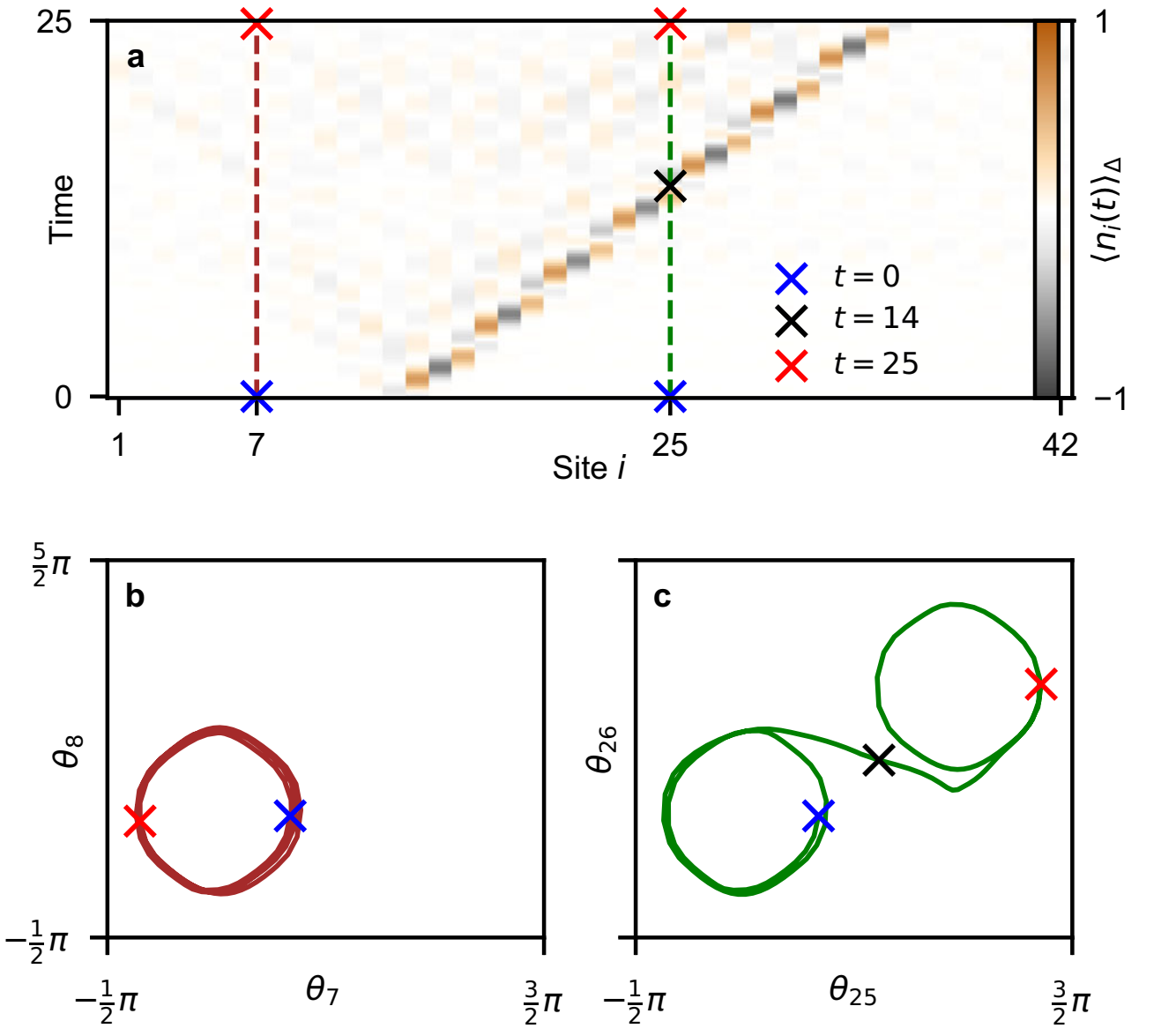

**Fig. 5 | Time evolution of classical equations of motion (9) of an infinite chain with a 42-site unit cell up to time $t = 25$.** The initial state has all sites initialized with $\theta^S_{1,2,3}$ angles except for the sites {13, 14, 15} initialized with right-moving soliton $\theta^R_{1,2,3}$ angles. **a** The difference in the expectation value of the number operator between states with and without a soliton shows the predominant propagation of the disturbance to the right. **b**, **c** Dynamics of $\theta_{7,8}$ and $\theta_{25,26}$ are constrained to approximately closed trajectories. The propagation of the soliton through sites 25 and 26 moves the dynamics of the corresponding angles to a different, almost closed orbit.

## Discussion

We have demonstrated that the PXP model in one-dimensional atom chains hosts discrete quasi-solitons—localized, potentially energy-carrying, wave packets that propagate coherently over long distances. The quasi-solitons identified in our work propagate on top of special reviving initial states, and do not have a dispersion, as their velocity is fixed by the unit cell size and the revival period of the underlying state. Quasi-solitons exhibit weak decay if propagating in the same direction, and stronger decay upon collisions. Importantly, they enable directed transport of energy and quantum information, suggesting potential uses in communication protocols in synthetic quantum systems. We have also shown that similar coherent structures emerge in a classical system governed by nonlinear equations of motion derived via TDVP.

We want to highlight how unusual the coherent transport of the quasi-solitons in the background of these special reviving states is. All states considered in our work correspond to highly-excited (if not infinite temperature) states of the system, and the coherent propagation of quasi-solitons is not a manifestation of low-energy physics. Furthermore, soliton cells $|L, R\rangle$ are orthogonal to the cells of the reviving state, $|S\rangle$, and therefore cannot be understood as a weak deformation thereof. Their spectral decomposition is also much less straightforward than for scarred states, whose dynamics can be explained via the anomalously high overlap with only a few non-thermal eigenstates[13]. The initial states with embedded quasi-solitons also exhibit several outliers, but they are much less distinctly separated from the bulk, meaning that many other eigenstates contribute meaningfully to the dynamics. We note that the outliers are different than those previously identified as causing revivals from the $|\mathbb{Z}_2\rangle$ states in the same model, but actually match well with those responsible for the oscillatory behavior of the $|K\rangle$ state that serves as a background for the solitons (see Supplementary Fig. 7[30]).

Although we have established the existence of quasi-solitons, they lack a complete theoretical description, and many aspects remain to be investigated. Inserting left-/right-moving quasi-solitons provides the PXP model with an exponentially large manifold of non-thermal multi-soliton states, but their relevance to physical properties of the model, such as its superdiffusive energy transport[15], remains to be understood, as well as the possible existence of other soliton-hosting states. Stabilizing quasi-solitons by deforming the PXP model[14,20,34,35] and thereby establishing the existence of fully stable solitons, applying driving[12,22] or potential confinement[36] would allow us to utilize the exponentially large non-thermal Hilbert space to store and transmit information, similar to what we have shown above. Beyond theoretical descriptions, experimental studies of quantum quenches from inhomogeneous multi-soliton initial states—feasible with current Rydberg array quantum simulators[10–12]—could provide important insights. In Supplementary Fig. 14[30], we provide additional evidence that our quasi-solitons can also be observed in experimentally realistic settings with long-range interactions and imperfect Rydberg blockade. A complementary angle to address these questions is via the study of the associated classical dynamical system (9), which hosts soliton-like excitations and may offer insights into quantum dynamics as well as connections to discrete solitons in optical and other systems[4].

Another direction is the possible connection to solitonic quasi-particles in integrable Floquet circuits related to the PXP model, such as the RCA201 cellular automaton[35]. This Floquet circuit can be obtained by discretizing (Trotterizing) the time-continuous PXP model and choosing a large step size. In principle, there is no physical argument why a connection between solitonic quasiparticles in the time-continuous model and the cellular automaton should exist, as the Trotter step size required to obtain the cellular automaton is beyond the perturbative regime. Nevertheless, ref. 34 found that certain quantum many-body scars of the time-continuous PXP model, associated with the $|\mathbb{Z}_2\rangle$ state, can be adiabatically connected to cellular automaton eigenstates by treating the Trotter step size as an interpolation parameter. Moreover, certain dynamical signatures of cellular automaton soliton product states persist down to small Trotterization step sizes. This raises the question of whether features of our quasi-solitons in the time-continuous chaotic PXP model are related to the dynamical features of simpler, analytically tractable Floquet realizations, such as the cellular automaton. However, our quasi-soliton states are weakly entangled, so a meaningful connection would require deforming not only the step size but also the state itself into a cellular automaton product state, and it is not obvious how to do so. A more detailed discussion and initial explorations are provided in Supplementary Note 5[30], leading to the conclusion that a connection is not apparent.

More generally, it would be interesting to understand if other scarred discrete lattice models can host solitonic excitations. In Supplementary Fig. 15[30], we show that the PPXPP model[29] exhibits excitations that are similar but less stable. Other potential candidates include the spin-1 XY model, where scars[37] and unconventional energy transport were also reported[38], as well as kinetically constrained models such as higher-spin generalizations of PXP[23] or lattice gauge theories[39–41]. Even more intriguing are questions of potential generalization of present solitons to two-dimensional lattices and the relation to solitons in Rydberg atomic gases[42,43], potentially coupled to light[44] in a regime of electromagnetically induced transparency[45]. Making progress in some of the above questions will lead to a better understanding of dynamics in excited quantum systems and realizations of soliton physics in quantum simulators.

# Methods

## Time-evolving block decimation algorithm

Unitary time-evolutions in this article have been conducted via the time-evolving block decimation (TEBD) algorithm[46,47] using the ITensor library[48,49]. The dynamics generated by Eq. (1) are obtained by using a symmetric fourth-order integrator[50,51] with the time step chosen to be $\Delta t = 0.02$. All simulations have been performed with open boundary conditions and a maximal bond dimension $\chi = 256$. Singular values below the threshold $10^{-12}$ were discarded.

All local observables were calculated at least 30 sites away from the boundaries of the chain to eliminate potential boundary effects for the considered time windows. The translation fidelity defined in Eq. (7) is a global quantity and consequently affected by the boundaries. Therefore, we only consider a finite region of interest $A$ by tracing out its complement $\bar{A}$ and calculating the translation fidelity as:

$$F_m(t) = \langle K^{\mathbf{l},\mathbf{r}}(t)|\mathrm{tr}_{\bar{A}}(|K^{\mathbf{l}-m,\mathbf{r}+m}\rangle\langle K^{\mathbf{l}-m,\mathbf{r}+m}|)|K^{\mathbf{l},\mathbf{r}}(t)\rangle. \tag{10}$$

Here, the reduced operator obtained after tracing out $\bar{A}$ is understood to be embedded back into the full Hilbert space by tensoring with the identity on $\bar{A}$. If the region $A$ is chosen to be the entire chain, this expression becomes equivalent to Eq. (7).

To ensure that the data produced via TEBD is reasonably faithful up to the reported times, we repeated the simulations with bond dimension $\chi = 128$ and compared the results. For quantities $O$ bounded by 1.0 in their reported units (number operator, energy density operator and energy-energy correlation operator expectation value, as well as translation fidelity), we have ensured that their estimates produced with different bond dimensions, denoted as $O_\chi$, agree. Specifically, we check that $|O_{256} - O_{128}| \leq 0.01$ for all times. Similarly, for the bipartite entanglement entropy, we required that the relative error is upper bounded by $|S_{256} - S_{128}|/S_{256} \leq 0.01$ for all cuts at all times.

Finally, we provide the explicit form of the initial states of the TEBD simulations shown in Figs. 1– 4: Figure 1 : The initial state used in the simulation is defined on 150 sites and is given by $|S\rangle^{\otimes 17}\left|R^+\right\rangle|S\rangle^{\otimes 4}\left|L^+\right\rangle|S\rangle^{\otimes 3}\left|R^+\right\rangle^{\otimes 2}|R^-\rangle|S\rangle^{\otimes 2}\left|R^+\right\rangle|S\rangle^{\otimes 18}$ where the different 3-site cells are defined in the main text. Figure 2: The state $\left|K^{\mathbf{l},\mathbf{r}}\right\rangle$ is defined on 150 sites with $\mathbf{l} = \varnothing$, $\mathbf{r} = \{23\}$.

Figure 3 : All $\left|K^{\mathbf{l},\mathbf{r}}\right\rangle$ states are defined on 120 sites. The states denoted as $(SRS)^k$ for $k = 0\ldots4$ correspond to $\mathbf{l} = \varnothing$ and $\mathbf{r} = \{11 + 3i|1 \leq i \leq k\}$. The states denoted as $R^k$ correspond to $\mathbf{l} = \varnothing$, $\mathbf{r} = \{17 + i|1 \leq i \leq k\}$ for $k = 1\ldots4$. The state $RSSL$ is given by $\mathbf{l} = \{22\}$, $\mathbf{r} = \{19\}$ and $LR$ is given by $\mathbf{l} = \{20\}$, $\mathbf{r} = \{21\}$.

Figure 4 : The initial state is defined on 102 sites and is given by $|S\rangle^{\otimes 16}\left(\left|L^+R^+\right\rangle + |L^-R^-\rangle\right)|S\rangle^{\otimes 16}/\sqrt{2}$.

## Time-dependent variational principle

A low-dimensional manifold able to capture the scarring from the $\left|\mathbb{Z}_2\right\rangle$ and $\left|\mathbb{Z}_3\right\rangle$ states has previously been proposed for the PXP model[23,24]. We are using the same $\chi = 2$ matrix-product state (MPS) Ansatz with a different parametrization of the on-site matrices:

$$A_j^{\downarrow} = \begin{pmatrix} \cos\theta_j & 0 \\ 1 & 0 \end{pmatrix}, \qquad A_j^{\uparrow} = \begin{pmatrix} 0 & e^{i\phi_j}\sin\theta_j \\ 0 & 0 \end{pmatrix}, \tag{11}$$

and, therefore, spans the same manifold of states. The states of this manifold are defined on an infinite chain with a repeating unit cell of size $N$ and take the form:

$$\left|\psi(\{\theta_j,\phi_j\})\right\rangle = \lim_{\ell\to\infty}\sum_{\sigma_1,\ldots,\sigma_\ell} \mathrm{Tr}\left(\prod_{i=1}^{\ell} A_i^{\sigma_i}(\theta_{((i-1)\,\mathrm{mod}\,N)+1},\phi_{((i-1)\,\mathrm{mod}\,N)+1})\right)|\sigma_1\cdots\sigma_\ell\rangle, \tag{12}$$

where $\sigma \in \{\uparrow,\downarrow\}$ labels the on-site basis.

Importantly, this Ansatz is able to represent the scarred state $|K\rangle$ and any combination of soliton defects $|L,R\rangle$ embedded in the state (including energy-carrying solitons $\left|L^\pm,R^\pm\right\rangle$). The projection of the unitary dynamics generated by the PXP Hamiltonian onto this variational manifold yields a system of coupled nonlinear first-order differential equations. Following the scheme in ref. 24, equations of motion (EOMs) for the $\theta$ and $\phi$ angles can be obtained from $\sum_j 2\mathrm{Im}\langle\partial_{\theta_j}\psi|\partial_{\phi_k}\psi\rangle\partial_t\theta_j = \partial_{\phi_k}\langle\psi|H|\psi\rangle$ and $\sum_k 2\,\mathrm{Im}\left\langle\partial_{\theta_j}\psi\,|\,\partial_{\phi_k}\psi\right\rangle\partial_t\phi_k = -\partial_{\theta_j}\langle\psi|H|\psi\rangle$ for all $j,k \in [1,N]$, where $|\psi\rangle$ is the MPS with an infinitely repeating unit cell of size $N$, therefore depending on $N$ variational angles $\theta$ and $N$ variational angles $\phi$.

We have analytically derived the EOMs for a unit cell size of $N = 1$ to 4 and conjectured its general form to be:

$$\partial_t\theta_1 = -\cos\theta_2\sin\phi_1 - \sin^3\theta_1\cot\theta_N\frac{\langle n_N\rangle}{\langle n_1\rangle}\sin\phi_N, \tag{13}$$

$$\begin{aligned}\partial_t\phi_1 = &\left[\sin^3\theta_2\frac{\langle n_1\rangle\cot\theta_1}{\langle n_2\rangle\cot\theta_2} - 2\cos\theta_2\cot2\theta_1\right]\cos\phi_1\\ &+\cos\theta_3\tan\theta_2\cos\phi_2 + \sin\theta_1\frac{\langle n_N\rangle\cot\theta_N}{\langle n_1\rangle\cot\theta_1}\cos\phi_N,\end{aligned} \tag{14}$$

with the expectation value of the number operator $\langle n_j\rangle$ on site $j$ given by:

$$\langle n_1\rangle = \sin^2\theta_1\frac{\sum_{l=0}^{N-1}\prod_{j=1}^{l}(-\sin^2\theta_{N-j+1})}{1-\prod_{j=1}^{N}(-\sin^2\theta_j)}. \tag{15}$$

The EOMs and expectation values for other angles can be simply obtained by a cyclic permutation of the indices as $j \to j+1$. These equations of motion match those independently derived in ref. 33, and their equivalence is shown in Supplementary Note 4[30].

The plane with $\phi_j = \pi/2$ forms a flow-invariant subspace where $\partial_t\phi_j = 0\ \forall j$, as follows from Eq. (14). This submanifold is able to express all our states $\left|K^{\mathbf{l},\mathbf{r}}\right\rangle$ and the EOMs simplify to the ones given in Eq. (9). The number operator $\langle n_j\rangle$ is still given by Eq. (15) as this quantity is independent of the phase information encoded in the $\phi_j$ angles. In order to search for periodic trajectories and solitons, we evolve the system of first-order differential equations (9) numerically via SciPy's Runge-Kutta RK45 integrator of order 5(4)[52,53]. The numerically optimized angles, as described in Supplementary Note 4[30], are given by:

$$\theta^S_{1,2,3} = (1.05977133, 1.46820419, 0.0000131027276),$$
$$\theta^R_{1,2,3} = (0.990853599, 0.454003269, 3.10125177).$$

In Fig. 5, the angles $\theta^S_{1,2,3}$ were used as a background, and a right-moving soliton was launched by initializing a particular three-site unit cell with angles $\theta^R_{1,2,3}$.

# Data availability

The raw data used to generate the figures are available at ref. 54.

# Code availability

The TEBD algorithm used in this work was implemented using the ITensor library[48,49] and the integration of the classical differential equations was performed via SciPy's Runge-Kutta RK45 integrator[52,53]. The code used in this study to produce the plots from the shared data is available at ref.[54].

# References

1. Nguyen, J. H. V., Dyke, P., Luo, D., Malomed, B. A. & Hulet, R. G. Collisions of matter-wave solitons. *Nat. Phys.* **10**, 918–922 (2014).

2. Grams, C. P. et al. Observation of chiral solitons in LiCuVO4. *Commun. Phys.* **5**, 37 (2022).
3. Ustinov, A. V. et al. Dynamics of Sine-Gordon solitons in the annular Josephson junction. *Phys. Rev. Lett.* **69**, 1815–1818 (1992).
4. Lederer, F. et al. Discrete solitons in optics. *Phys. Rep.* **463**, 1–126 (2008).
5. Zabrodin, A. Classical facets of quantum integrability. Preprint at https://arxiv.org/abs/2501.18557 (2025).
6. Vlijm, R. et al. Quasi-soliton scattering in quantum spin chains. *Phys. Rev. B* **92**, 214427 (2015).
7. Fukuhara, T. et al. Microscopic observation of magnon bound states and their dynamics. *Nature* **502**, 76–79 (2013).
8. Fendley, P., Sengupta, K. & Sachdev, S. Competing density-wave orders in a one-dimensional hard-boson model. *Phys. Rev. B* **69**, 075106 (2004).
9. Lesanovsky, I. & Katsura, H. Interacting Fibonacci anyons in a Rydberg gas. *Phys. Rev. A* **86**, 041601(R) (2012).
10. Browaeys, A. & Lahaye, T. Many-body physics with individually controlled Rydberg atoms. *Nat. Phys.* **16**, 132–142 (2020).
11. Bernien, H. et al. Probing many-body dynamics on a 51-atom quantum simulator. *Nature* **551**, 579–584 (2017).
12. Bluvstein, D. et al. Controlling quantum many-body dynamics in driven Rydberg atom arrays. *Science* **371**, 1355–1359 (2021).
13. Turner, C. J., Michailidis, A. A., Abanin, D. A., Serbyn, M. & Papić, Z. Weak ergodicity breaking from quantum many-body scars. *Nat. Phys.* **14**, 745–749 (2018).
14. Choi, S. et al. Emergent SU(2) dynamics and perfect quantum many-body scars. *Phys. Rev. Lett.* **122**, 220603 (2019).
15. Ljubotina, M., Desaules, J.-Y., Serbyn, M. & Papić, Z. Superdiffusive energy transport in kinetically constrained models. *Phys. Rev. X* **13**, 011033 (2023).
16. Serbyn, M., Abanin, D. A. & Papić, Z. Quantum many-body scars and weak breaking of ergodicity. *Nat. Phys.* **17**, 675–685 (2021).
17. Moudgalya, S., Bernevig, B. A. & Regnault, N. Quantum many-body scars and Hilbert space fragmentation: A review of exact results. *Rep. Prog. Phys.* **85**, 086501 (2022).
18. Chandran, A., Iadecola, T., Khemani, V. & Moessner, R. Quantum many-body scars: A quasiparticle perspective. *Annu. Rev. Condens. Matter Phys.* **14**, 443–469 (2023).
19. Turner, C. J., Michailidis, A. A., Abanin, D. A., Serbyn, M. & Papić, Z. Quantum scarred eigenstates in a Rydberg atom chain: entanglement, breakdown of thermalization, and stability to perturbations. *Phys. Rev. B* **98**, 155134 (2018).
20. Khemani, V., Laumann, C. R. & Chandran, A. Signatures of integrability in the dynamics of Rydberg-blockaded chains. *Phys. Rev. B* **99**, 161101 (2019).
21. Park, H. K. & Lee, S. Graph-theoretical proof of nonintegrability in quantum many-body systems: Application to the PXP model. *Phys. Rev. B* **111**, L081101 (2025).
22. Maskara, N. et al. Discrete time-crystalline order enabled by quantum many-body scars: Entanglement steering via periodic driving. *Phys. Rev. Lett.* **127**, 090602 (2021).
23. Ho, W. W., Choi, S., Pichler, H. & Lukin, M. D. Periodic orbits, entanglement, and quantum many-body scars in constrained models: Matrix product state approach. *Phys. Rev. Lett.* **122**, 040603 (2019).
24. Michailidis, A. A., Turner, C. J., Papić, Z., Abanin, D. A. & Serbyn, M. Slow quantum thermalization and many-body revivals from mixed phase space. *Phys. Rev. X* **10**, 011055 (2020).
25. Surace, F. M. et al. Lattice gauge theories and string dynamics in Rydberg atom quantum simulators. *Phys. Rev. X* **10**, 021041 (2020).
26. Chen, G., Huang, W. & Yao, Y. Dynamics of entangled domain walls in the PXP model under driving: Crossover from prethermalization to localization. *Phys. Rev. B* **106**, 174302 (2022).
27. Desaules, J.-Y. et al. Ergodicity breaking under confinement in cold-atom quantum simulators. *Quantum* **8**, 1274 (2024).
28. Su, G.-X., Osborne, J. J. & Halimeh, J. C. Cold-atom particle collider. *PRX Quantum* **5**, 040310 (2024).
29. Kerschbaumer, A., Ljubotina, M., Serbyn, M. & Desaules, J.-Y. Quantum many-body scars beyond the PXP model in Rydberg simulators. *Phys. Rev. Lett.* **134**, 160401 (2025).
30. Supplementary Online Material (SOM). The SOM contains additional data and analyses that support the results in the main text. Refs. [12, 13, 16-18, 24, 29, 33-35, 52, 55, 56] are used in the SOM.
31. Dooley, S., Johnston, L., Gormley, P. & Campbell, B. Perfect quantum state transfer through a chaotic spin chain via many-body scars. *Phys. Rev. B* **112**, 214304 (2025).
32. Cesa, F. & Pichler, H. Universal Quantum Computation in Globally Driven Rydberg Atom Arrays. *Phys. Rev. Lett.* **131**, 170601 (2023).
33. Hu, Z. & Wu, B. Variational method for $\mathbb{Z}_K$ wavefunctions in spin-$J$ PXP model. Preprint at https://arxiv.org/abs/2501.09301 (2025).
34. Giudici, G., Surace, F. M. & Pichler, H. Unraveling PXP many-body scars through Floquet dynamics. *Phys. Rev. Lett.* **133**, 190404 (2024).
35. Wilkinson, J. W. P., Klobas, K., Prosen, T. & Garrahan, J. P. Exact solution of the Floquet-PXP cellular automaton. *Phys. Rev. E* **102**, 062107 (2020).
36. Kormos, M., Collura, M., Takács, G. & Calabrese, P. Real-time confinement following a quantum quench to a non-integrable model. *Nat. Phys.* **13**, 246–249 (2017).
37. Schecter, M. & Iadecola, T. Weak ergodicity breaking and quantum many-body scars in spin-1 XY magnets. *Phys. Rev. Lett.* **123**, 147201 (2019).
38. Morettini, G., Capizzi, L., Fagotti, M. & Mazza, L. Transport in a system with a tower of quantum many-body scars. *Phys. Rev. B* **112**, 134314 (2025).
39. Chandrasekharan, S. & Wiese, U.-J. Quantum link models: a discrete approach to gauge theories. *Nucl. Phys. B* **492**, 455–471 (1997).
40. Wiese, U.-J. Ultracold quantum gases and lattice systems: quantum simulation of lattice gauge theories. *Ann. der Phys.* **525**, 777–796 (2013).
41. Hauke, P., Marcos, D., Dalmonte, M. & Zoller, P. Quantum simulation of a lattice Schwinger model in a chain of trapped ions. *Phys. Rev. X* **3**, 041018 (2013).
42. Maucher, F. et al. Rydberg-induced solitons: three-dimensional self-trapping of matter waves. *Phys. Rev. Lett.* **106**, 170401 (2011).
43. Zhao, Y. et al. Three-dimensional solitons in Rydberg-dressed cold atomic gases with spin-orbit coupling. *Sci. Rep.* **13**, 18079 (2023).
44. Bai, Z., Zhang, Q. & Huang, G. Quantum reflections of nonlocal optical solitons in a cold Rydberg atomic gas. *Phys. Rev. A* **101**, 053845 (2020).
45. Fleischhauer, M., Imamoglu, A. & Marangos, J. P. Electromagnetically induced transparency: Optics in coherent media. *Rev. Mod. Phys.* **77**, 633–673 (2005).
46. Vidal, G. Classical simulation of infinite-size quantum lattice systems in one spatial dimension. *Phys. Rev. Lett.* **98**, 070201 (2007).
47. Schollwöck, U. The density-matrix renormalization group in the age of matrix product states. *Ann. Phys.* **326**, 96 – 192 (2011).
48. Fishman, M., White, S. R. & Stoudenmire, E. M. The ITensor Software Library for Tensor Network Calculations. *SciPost Phys. Codebases* 4 (2022).
49. Fishman, M., White, S. R. & Stoudenmire, E. M. Codebase release 0.3 for ITensor. *SciPost Phys. Codebases* 4–r0.3 (2022).
50. Forest, E. & Ruth, R. D. Fourth-order symplectic integration. *Phys. D: Nonlinear Phenom.* **43**, 105–117 (1990).
51. Yoshida, H. Construction of higher order symplectic integrators. *Phys. Lett. A* **150**, 262–268 (1990).
52. Virtanen, P. et al. SciPy 1.0: Fundamental algorithms for scientific computing in Python. *Nat. Methods* **17**, 261–272 (2020).

53. Dormand, J. & Prince, P. A family of embedded Runge-Kutta formulae. *J. Comput. Appl. Math.* **6**, 19–26 (1980).
54. Kerschbaumer, A. Research Data: Quasi-solitons in Rydberg atom chains. Institute of Science and Technology Austria https://doi.org/10.15479/AT-ISTA-21960 (2026).

## Acknowledgements
We acknowledge useful discussions with J.-S. Caux, E. Demler, J. Dubail, F. Essler, J. Feldmeier, S. Garratt, W. W. Ho, M. Lukin, Z. Papic, S. Rotter, F. Surace, and R. Vasseur.

## Author contributions
A.K., J.-Y.D., M.L. and M.S. performed numerical and analytical calculations, generated ideas, contributed to discussions, and worked on the manuscript. A.K. prepared the bulk of the manuscript. M.S. conceived and supervised the project.

## Funding
J.-Y.D. acknowledges funding from the European Union's Horizon 2020 research and innovation programme under the Marie Skłodowska-Curie Grant Agreement No. 101034413. M. L. acknowledges support by the Deutsche Forschungsgemeinschaft (DFG, German Research Foundation) under Germany's Excellence Strategy—EXC-2111—390814868. We acknowledge support by the Erwin Schrödinger International Institute for Mathematics and Physics (ESI). This research was funded in part by the Austrian Science Fund (FWF) https://doi.org/10.55776/COE1 and the European Union—NextGenerationEU. This research was supported in part by grant NSF PHY2309135 to the Kavli Institute for Theoretical Physics (KITP).

## Competing interests
The authors declare no competing interests.

## Additional information
**Supplementary information** The online version contains supplementary material available at https://doi.org/10.1038/s41467-026-75598-1.

**Correspondence** and requests for materials should be addressed to Aron Kerschbaumer, Jean-Yves Desaules, Marko Ljubotina or Maksym Serbyn.

**Peer review information** *Nature Communications* thanks the anonymous reviewers for their contribution to the peer review of this work. A peer review file is available

**Reprints and permissions information** is available at http://www.nature.com/reprints

# Supplementary Information for "Quasi-solitons in Rydberg atom chains"

Aron Kerschbaumer [1], Jean-Yves Desaules [1], Marko Ljubotina [1,2,3], and Maksym Serbyn [1]
[1] Institute of Science and Technology Austria (ISTA), Am Campus 1, 3400 Klosterneuburg, Austria
[2] Physics Department, Technical University of Munich, TUM School of Natural Sciences, James-Franck-Str. 1, Garching 85748, Germany
[3] Munich Center for Quantum Science and Technology (MCQST), Schellingstr. 4, München 80799, Germany

## CONTENTS

## Supplementary Note 1. SOLITONS UNDER UNITARY DYNAMICS

In Fig. 2**d** of the main text, we have shown the approximate dynamics, during one revival period $T$, of the scarred initial state $|K\rangle$ and the same state perturbed with a single right-moving cell $|R\rangle$. We validate this picture by numerically simulating a small system with $N = 15$ sites with periodic boundary conditions. We calculate the overlaps of the time-evolved states with the Ansatz states that obey the same structure as proclaimed in the main text figure. The relatively modest system size was chosen to focus on the local dynamics of the soliton and not the scarred cells $|S\rangle$ far away from the soliton and their decay due to imperfect oscillations. The periodic boundary conditions are chosen to avoid the soliton being disturbed by boundary effects. To investigate the structure of $|K(t)\rangle$ we introduce the states:

$$|\psi^0\rangle = |S\rangle^{\otimes 5}, \quad |\psi^{T/4}\rangle = \mathcal{T}\,|R\rangle^{\otimes 5}, \quad |\psi^{T/2}\rangle = \mathcal{T}^2\,|\bar{S}\rangle^{\otimes 5}, \quad |\psi^{3T/4}\rangle = \mathcal{T}\,|L\rangle^{\otimes 5}, \tag{1}$$

where $\mathcal{T}$ is the translation operator that transforms each site $i \to i+1$. The cell $|\bar{S}\rangle$, depicted in yellow in Fig. 2**d** is given by $|\bar{S}\rangle = \frac{1}{\beta}|\downarrow\rangle(|\uparrow\downarrow\rangle + \beta|\downarrow\uparrow\rangle)$. Supplementary Fig. 1**a** shows that the state $|K(mT/4)\rangle$ indeed displays large overlap with the Ansatz state $|\psi^{mT/4}\rangle$ for all integer $m$, with the corresponding times $mT/4$ highlighted by vertically dashed lines.

For the state with a single right-moving soliton $K^{\mathbf{l},\mathbf{r}}$ with $\mathbf{r} = \{1\}$ and $\mathbf{l} = \varnothing$, we use a similar approach and define the following states:

$$|\psi_S^0\rangle = |R\rangle\,|S\rangle^{\otimes 4}, \quad |\psi_S^{T/4}\rangle = \mathcal{T}\,|\bar{S}\rangle\,|R\rangle^{\otimes 4}, \quad |\psi_S^{T/2}\rangle = \mathcal{T}^2\,|D^1\rangle\,|\bar{S}\rangle^{\otimes 3}, \quad |\psi_S^{3T/4}\rangle = \mathcal{T}\,|L\rangle^{\otimes 2}\,|D^2\rangle\,|L\rangle, \tag{2}$$

where the 6-site defect cells at $T/2$ and $3T/4$ are given by:

$$\begin{aligned}
|D^1\rangle = d_1\,|\downarrow\rangle\Big(&0.410\,|\downarrow\uparrow\downarrow\downarrow\downarrow\rangle - 0.342\,\mathrm{i}\,|\downarrow\uparrow\downarrow\downarrow\uparrow\rangle + 0.176\,\mathrm{i}\,|\downarrow\uparrow\downarrow\uparrow\downarrow\rangle + |\uparrow\downarrow\downarrow\downarrow\downarrow\rangle \\
&- 0.834\,\mathrm{i}\,|\uparrow\downarrow\downarrow\downarrow\uparrow\rangle + 0.247\,\mathrm{i}\,|\uparrow\downarrow\downarrow\uparrow\downarrow\rangle + 0.387\,\mathrm{i}\,|\uparrow\downarrow\uparrow\downarrow\downarrow\rangle + 0.323\,|\uparrow\downarrow\uparrow\downarrow\uparrow\rangle\Big), \\
|D^2\rangle = d_2\,|\downarrow\rangle\Big(&0.136\,|\downarrow\downarrow\downarrow\downarrow\downarrow\rangle + 0.084\,\mathrm{i}\,|\downarrow\downarrow\downarrow\downarrow\uparrow\rangle + 0.139\,\mathrm{i}\,|\downarrow\downarrow\downarrow\uparrow\downarrow\rangle + \mathrm{i}\,|\downarrow\downarrow\uparrow\downarrow\downarrow\rangle \\
&- 0.614\,|\downarrow\downarrow\uparrow\downarrow\uparrow\rangle + 0.113\,\mathrm{i}\,|\downarrow\uparrow\downarrow\downarrow\downarrow\rangle - 0.069\,|\downarrow\uparrow\downarrow\downarrow\uparrow\rangle + 0.132\,|\downarrow\uparrow\downarrow\uparrow\downarrow\rangle \\
&+ 0.071\,|\uparrow\downarrow\downarrow\uparrow\downarrow\rangle + 0.389\,|\uparrow\downarrow\uparrow\downarrow\downarrow\rangle + 0.239\,\mathrm{i}\,|\uparrow\downarrow\uparrow\downarrow\uparrow\rangle\Big),
\end{aligned} \tag{3}$$

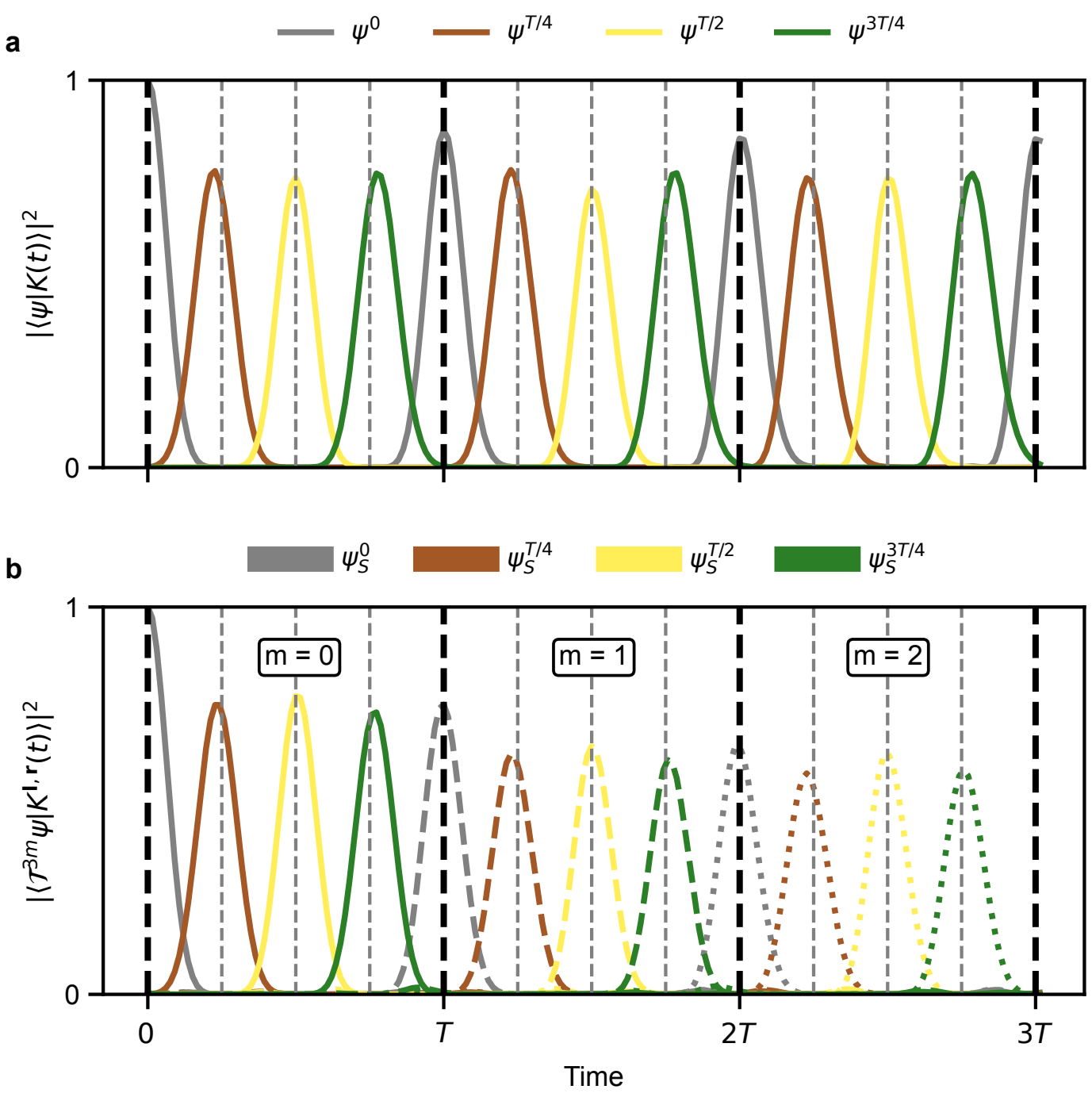

Supplementary Fig. 1. Exact diagonalization simulation of the dynamics of a scarred state without and with right-moving defect cell. The system simulated is a chain with 15 sites with periodic boundary conditions, for which the first three revival periods are shown. Vertical dashed lines indicate times multiple of $T/4$. **a,** The fidelity between $|K(t)\rangle$ and the Ansatz states defined in Eq. (1) show large peaks at multiples of $T/4$. **b,** Similar peaks are visible in the fidelity between the state $|K^{\mathbf{l},\mathbf{r}(t)}(t)\rangle$ ($\mathbf{l}(t) = \varnothing$ $\mathbf{r} = \{1\}$) and the Ansatz states defined in Eq. (2), as well as with the Ansatz states translated by one unit cell, $\mathcal{T}^3$, (dashed lines) and two unit cells, $\mathcal{T}^6$ (dotted lines).

with $d_1$ and $d_2$ being normalization constants. The defect cells have been obtained by variationally optimizing generic 6-site states in order to maximize the overlaps between $|\psi_S^{T/2}\rangle$, $|\psi_S^{3T/4}\rangle$ and $K^{\mathbf{l},\mathbf{r}}$ at $T/2$ and $3T/4$ respectively. Even though it is unlikely that we have found the ideal defect cells that yield the maximal overlaps, they already reproduce the proclaimed soliton dynamics. This is visible in Supplementary Fig. 1**b** as pronounced fidelity peaks of the time-evolved state $K^{\mathbf{l},\mathbf{r}(t)}$ and the Ansatz states during the first revival period. Since the $R$ cell shifts by 3 sites to the right during each period $T$, we also evaluate the overlaps with $\mathcal{T}^3$ as well as $\mathcal{T}^6$ translated Ansatz states (colored dashed and dotted lines respectively).

The coherent propagation of the right-moving soliton, together with the chiral symmetry of the Hamiltonian in the form of $\{H, \mathcal{C}\} = 0$ and the enforced condition on the left-moving soliton $|L\rangle = \mathcal{C}\,|R\rangle$, ensures the approximate leftwards propagation of $|L\rangle$. Noting that the scarred cell satisfies $\mathcal{C}\,|S\rangle = |S\rangle$ it follows that $\mathcal{C}K^{\mathbf{l},\mathbf{r}} = K^{\mathbf{r},\mathbf{l}}$, so right-moving soliton cells are mapped to left-moving ones and vice versa. Let us consider a state with a single defect cell $|L\rangle$ at position $j$, $\mathbf{l} = \{j\}$, $\mathbf{r} = \varnothing$. Then the time evolution of this state for one period $T$ yields indeed:

$$e^{-iHT}\,|K^{\{j\},\varnothing}\rangle = e^{-iHT}\mathcal{C}\,|K^{\varnothing,\{j\}}\rangle = \mathcal{C}e^{+iHT}\,|K^{\varnothing,\{j\}}\rangle \approx \mathcal{C}\,|K^{\varnothing,\{j-1\}}\rangle = |K^{\{j-1\},\varnothing}\rangle\,, \tag{4}$$

where the step from the second to the third line relies on the approximate translation of $|R\rangle$ by one unit cell during one period $T$. Therefore, a coherently right-propagating soliton yields automatically a coherently left-propagating soliton.

That the left-moving soliton cell $|L\rangle$ embedded in the scarred background $|K\rangle$ shows indeed the same coherent propagation as $|R\rangle$ but in the opposite direction can be seen in Supplementary Fig. 2**a** in terms of the number operator and translation fidelity revivals. In the main text, we demonstrated that not only can a single soliton propagate coherently, but a bulk of solitons do as well. We can extend this bulk and consider a state comprising only $|L\rangle$ cells except a single $|S\rangle$ cell. All $|L\rangle$ cells move left by three sites per period; therefore, the single $|S\rangle$ cell is carried along as a defect in the same direction (panel **b** of Supplementary Fig. 2). The heightened translation fidelity revivals are being attributed to the slower decay of the scarred cells $|L\rangle$ compared to $|S\rangle$. However, we emphasize that the defect cells $|L, R\rangle$ only show coherent propagation if they are embedded in the corresponding scarred background. Embedded in a thermal state, they instead lead to incoherent propagation in both directions of the chain. This can be seen in Supplementary Fig. 2**c** for the cell $|L\rangle$ inserted in the polarized state $|0\rangle = |{\downarrow\downarrow}\ldots{\downarrow}\rangle$. While the bidirectional propagation of the defect is clearly visible from the number operator results, its incoherent nature is attested by the lack of peaks in the translation fidelity.

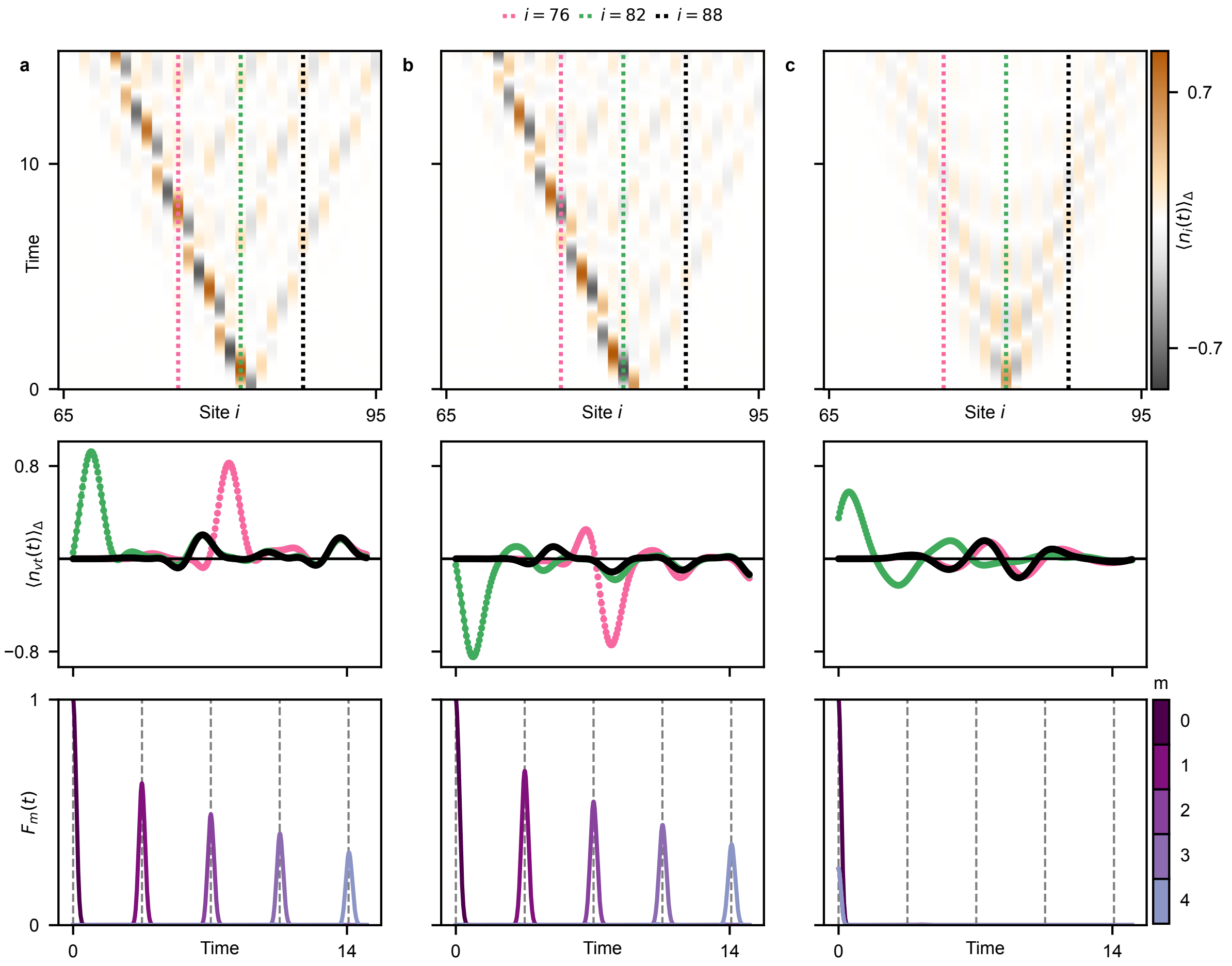

Supplementary Fig. 2. Dynamics of homogeneous initial states with 150 sites perturbed by a single defect cell positioned at sites $\{82, 83, 84\}$. The top row shows the difference between the number operators of the states with and without defect cells. The second row shows the same number operator difference over time for 3 selected sites corresponding to the vertical dashed lines in the top row. The bottom row shows translation fidelities of the corresponding states measured on the subsystem $A = \{61, 62, \ldots, 90\}$. **a,** Scarred state $|K\rangle$ with a defect cell $|L\rangle$ shows coherent propagation of left-moving defect. **b,** The initial state consisting only of left-moving cells $|L\rangle$ except the defect cell $|S\rangle$ show a similar leftwards propagation of the defect cell. **c,** Polarized state consisting of unit cells $|{\downarrow\downarrow\downarrow}\rangle$ where a defect cell $|L\rangle$ is inserted. The defect spreads incoherently in both directions of the chain, and no translation fidelity peak can be observed. In panels **c**, the definition of the translation fidelity is adapted by using $|{\downarrow\downarrow\downarrow}\rangle$ instead of the scarred background state $|S\rangle$.

Since the states consisting of the unit cells $|S\rangle$, $|L\rangle$ and $|R\rangle$ lie on the same scarred trajectory (up to translation), the homogeneous states $|K_L\rangle = |L\rangle^{\otimes m}$ and $|K_R\rangle = |R\rangle^{\otimes m}$ show periodic revivals as well, as shown in Supplementary Fig. 3. In comparison, the polarized state $|0\rangle$ shows rapidly decaying fidelity. The state $|K_{\bar{S}}\rangle = |\bar{S}\rangle^{\otimes m}$ shows the same scarred dynamics as $|S\rangle$, $|L\rangle$ and $|R\rangle$ but is omitted for clarity. In fact, we can understand its dynamics and the effect of inserting defects on top of it directly from the symmetries of the Hamiltonian and the structure of the various states. The spatial inversion operator I maps site $i \to N - i + 1$ and commutes with the PXP Hamiltonian. Assuming for simplicity periodic boundary conditions, one can see that the scarred cells $|S\rangle$ and $|\bar{S}\rangle$ are related by $|\bar{S}\rangle = \mathcal{T}\mathcal{I}\,|S\rangle$. Furthermore, the spatially inverted right-moving soliton cell has high overlap with the left-moving cell, up to translation. More specifically, $|\langle L|\mathcal{T}\mathcal{I}|R\rangle|^2 \approx 0.92$. We can define $K^{\mathbf{l},\mathbf{r}}_{|\bar{S}\rangle}$ analogously to Eq. (5) of the main text but formed of unit cells $|\bar{S}\rangle$ instead of $|S\rangle$, while $\mathbf{l}, \mathbf{r}$ again denote the positions of the defect cells $|L, R\rangle$. Considering such a state with an embedded right-moving defect cell $|R\rangle$ at position $j$, (i.e. $\mathbf{l} = \varnothing$, $\mathbf{r} = \{j\}$), it will evolve as:

$$e^{-iHT}\,|K_{\bar{S}}^{\varnothing,\{j\}}\rangle = \mathcal{I}e^{-iHT}\mathcal{I}\,|K_{\bar{S}}^{\varnothing,\{j\}}\rangle \approx \mathcal{I}e^{-iHT}\mathcal{T}^{-1}\,|K^{\{N-j+1\},\varnothing}\rangle \approx \mathcal{I}\mathcal{T}^{-1}\,|K^{\{N-j\},\varnothing}\rangle = |K_{\bar{S}}^{\varnothing,\{j+1\}}\rangle\,. \tag{5}$$

As such, $|L\rangle$ and $|R\rangle$ cells placed on top of the scarred state consisting of $|\bar{S}\rangle$ unit cells propagate to the left and right, respectively, as is the case for a background of $|S\rangle$ cells.

In summary, all unit cells attained at multiples of $T/4$ on the scarred trajectory can host coherently propagating defect cells. For $|K\rangle$ and $|K_{\bar{S}}\rangle$, inserting a $|R\rangle$ cell leads to propagation to the right, while inserting $|L\rangle$ leads to propagation to the left. Notably, for $|K_R\rangle$, only right-moving defects can be created by either inserting $|S\rangle$ or $|\bar{S}\rangle$, while the same defects lead to propagation to the left when inserted in $|K_L\rangle$.

A characteristic property of scarred states is their slow growth of entanglement under unitary time-evolution compared to thermal states. While their entanglement entropy still grows linearly in time, as shown in previous works [1, 2], this growth is strongly suppressed. Inserting soliton defect cells into the scarred background does not enhance the entanglement growth of the scarred state significantly, as can be seen in Supplementary Fig. 4. The evolution of the half-chain bipartite entanglement entropy $S_{\text{ent}} = -\,\text{Tr}\,[\rho_A \ln \rho_A]$ is being measured for different states $K_{|\bar{S}\rangle}^{\mathbf{l},\mathbf{r}}$ and all of them show oscillatory dynamics that is connected to reviving of the scarred cells, while a heightened entanglement peak occurs when the solitons pass through the observed cut. Moreover, these states show strongly suppressed entanglement growth compared to the thermal polarized state $|0\rangle$. Even the state showing a collision between left- and right-moving solitons $|\ldots SRSSLS\ldots\rangle$ still features a suppressed entanglement growth.

## Supplementary Note 2. ENERGY-CARRYING SOLITONS AND NEW SCARRED EIGENSTATES

Energy-carrying solitons $|R^{\pm}\rangle$ and $|L^{\pm}\rangle$ embedded in the scarred state $|K\rangle$ propagate coherently and show increased stability compared to their counterparts with zero energy, as shown in the main text. To better understand the interaction between energy-solitons and their decay, we are showing the local energy density of different multi-soliton configurations in Supplementary Fig. 5 and plot the energy densities along three different world-lines. As expected, the energy density along a soliton trajectory without collision decays very slowly, while the energy density experiences an additional decay during collision.

Energy-carrying soliton cells $|R^{\alpha}\rangle$ can be used to create homogeneous scarred states with tunable energy of the form $|K_{\alpha}\rangle = |R^{\alpha}\rangle^{\otimes l}$. The state returns approximately to its initial configuration because each cell shifts 3 sites to the right per period (scarred states can be formed analogously by using $|L^{\alpha}\rangle$). The energies at each site are given by:

$$\langle R^{\alpha}|X_1|R^{\alpha}\rangle = 0, \quad \langle R^{\alpha}|X_2|R^{\alpha}\rangle = \sqrt{\tfrac{1}{2+2/\beta^2}}\,\sin(2\alpha)\,, \quad \langle R^{\alpha}|X_3|R^{\alpha}\rangle = \sqrt{\tfrac{1}{2+2\beta^2}}\,\sin(2\alpha)\,, \tag{6}$$

leading to average energy per site $\langle\bar{X}\rangle$ between $-0.33$ and $0.33$ for the value of $\beta = 0.65$ that we use. For the ground and ceiling states, the same quantity takes values of $\pm 0.60$ respectively. Thus, the states $|R^{\alpha}\rangle$ and $|L^{\alpha}\rangle$ are always at

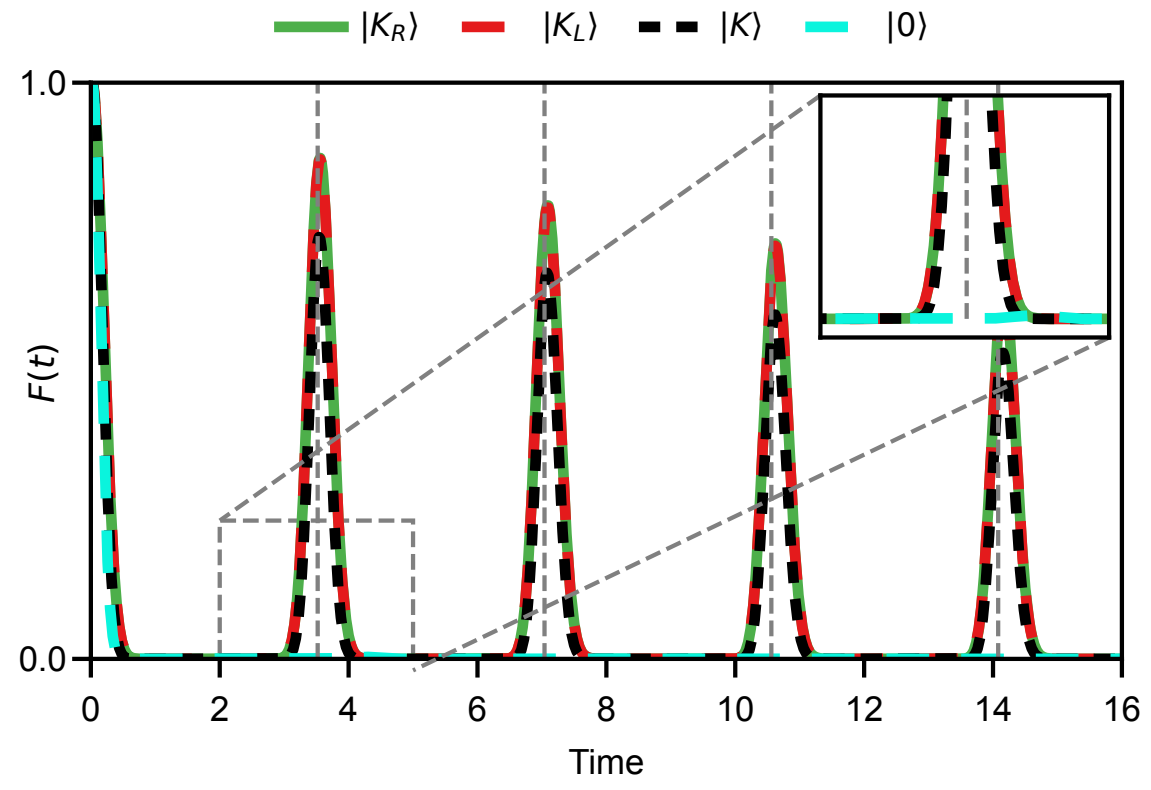

Supplementary Fig. 3. Time-evolution of different homogeneous initial states defined on a 150-site Rydberg chain. The states $|K_R\rangle = |R\rangle^{\otimes 50}$, $|K_L\rangle = |L\rangle^{\otimes 50}$ and $|K\rangle = |S\rangle^{\otimes 50}$ show scarred dynamics in form of large fidelity revivals at multiples of period $T$ (identified by vertically dashed lines). By comparison the polarized state $|0\rangle = |{\downarrow}\rangle^{\otimes 150}$ shows fast fidelity decay without revivals. Fidelities are being measured on the subsystem $A = \{61, 62, \ldots 90\}$.

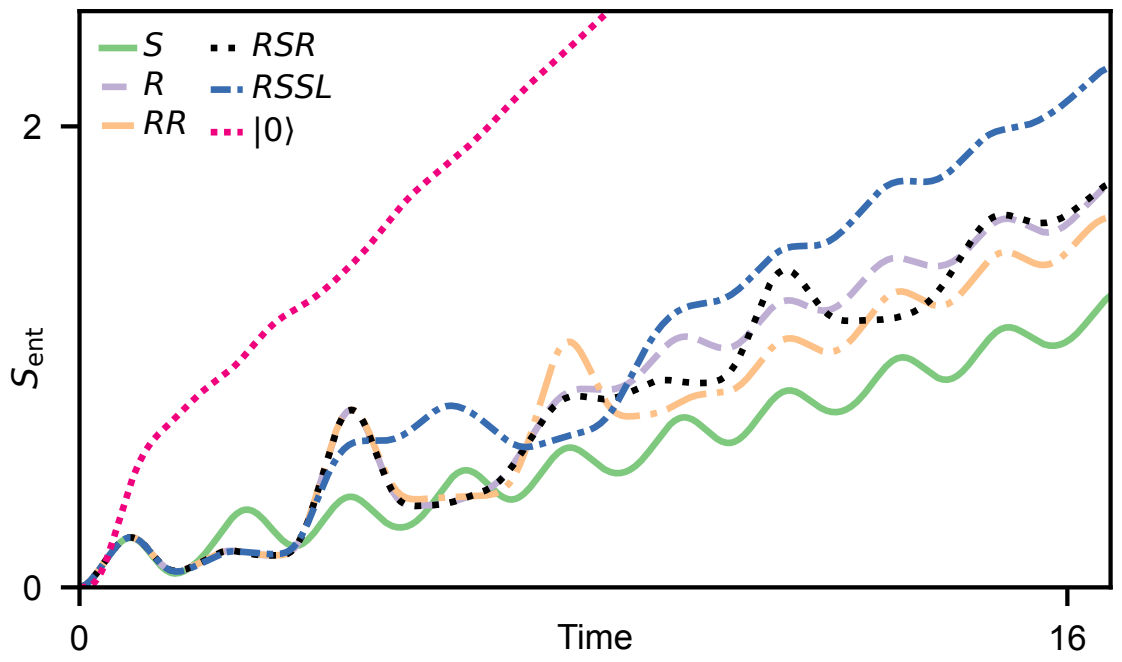

Supplementary Fig. 4. Half-chain bipartite entanglement dynamics of different $|K^{\mathbf{l},\mathbf{r}}\rangle$ states defined on a 120-site chain with the bipartition cut chosen between sites 60 and 61. The states show oscillatory entanglement dynamics, with additional peaks when the soliton passes through the cut, except for the state $RSSL$, which leads to a disruption of the oscillatory entanglement growth. The entanglement growth of all $|K^{\mathbf{l},\mathbf{r}}\rangle$ states is suppressed compared to the polarized state $|0\rangle$. The label $S$ corresponds to the scarred states $|K\rangle$. Label $R$: $\mathbf{l} = \varnothing$ and $\mathbf{r} = \{19\}$. Label $RR$: $\mathbf{l} = \varnothing$ and $\mathbf{r} = \{18, 19\}$, Label $RSR$: $\mathbf{l} = \varnothing$ and $\mathbf{r} = \{17, 19\}$, Label $RSSL$: $\mathbf{l} = \{22\}$ and $\mathbf{r} = \{19\}$

.

finite energy density away from the edges of the spectrum; therefore, excluding low-energy phenomena as a possible explanation for their revivals. Periodic revivals for states with different energy can be seen in Supplementary Fig. 6. While the fidelity decays faster than for zero-energy states, there are still strong signatures of ergodicity breaking for *all* values of $\alpha$.

## Supplementary Note 3. SPECTRAL DECOMPOSITION OF SOLITONS

Periodic revivals of scarred states are usually attributed to large overlaps with a set of special, equidistantly spaced, eigenstates [1, 3–5]. Indeed, in agreement with this intuition, the zero-energy state $|K\rangle$ shows high overlap with a band of special eigenstates that is symmetrically distributed around the center of the spectrum, as visible in Supplementary Fig. 7**a**. Likewise, the scarred state $|K_{\pi/4}\rangle$ built from repeating soliton cells $|R^{+}\rangle$ has enhanced overlaps with a band of special eigenstates that now all have positive energies, see Supplementary Fig. 7**b**. We emphasize that these eigenstates are fully distinct from the ones responsible for the scarring observed in the $|\mathbb{Z}_2\rangle$ state. Additionally, we note that these eigenstates would be hard to identify from entanglement entropy alone, as their entropies increase progressively as their energy approaches zero.

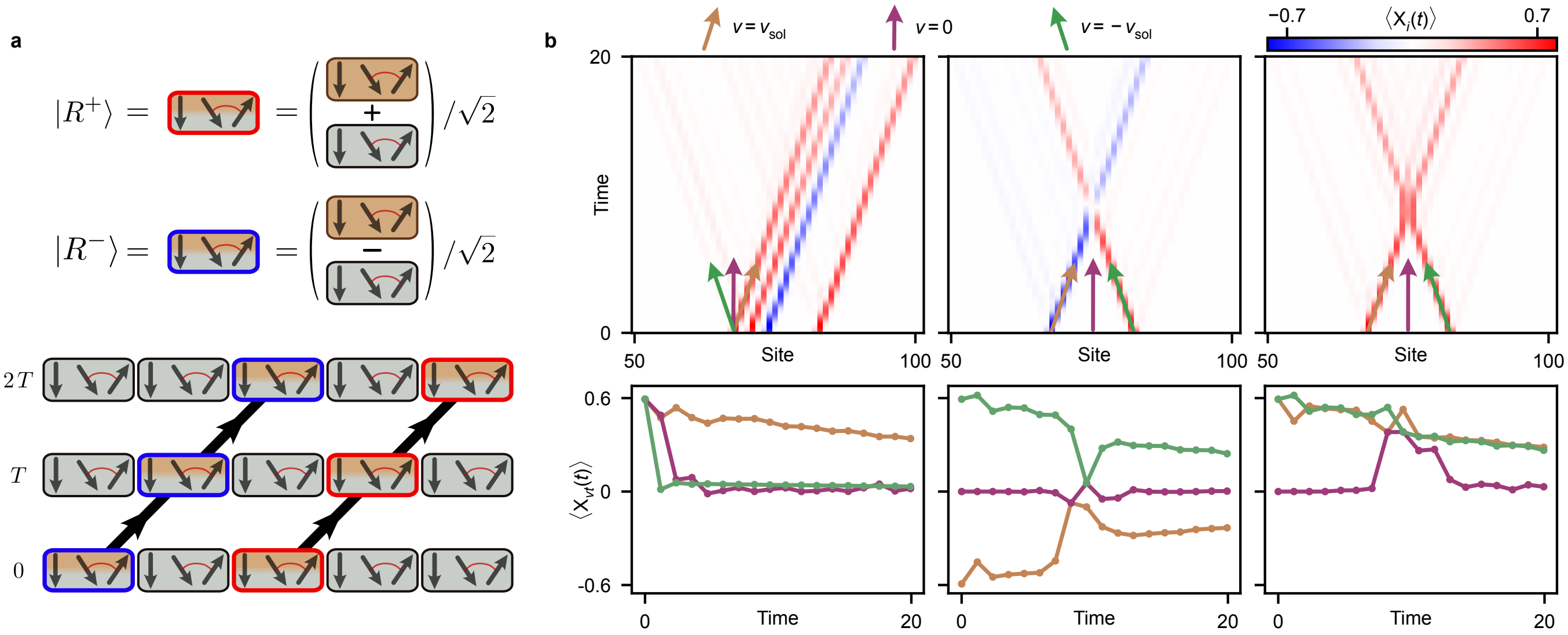

Supplementary Fig. 5. Time evolution of the energy density for three different energy-carrying multi-soliton states. Top: energy density dynamics shown as a function of position and time highlight the chiral propagation of energy. Bottom: energy density along world-lines with velocities $v = 0, \pm v_{\rm sol}$, with $v_{\rm sol} = 3/T \approx 0.85$. Without collisions (**a**), more than 50% of the original energy is retained in the soliton even at late times, whereas collisions (**b** and **c**) lead to an additional decrease in the energy density along the soliton trajectory. All three different initial states are defined on 150 sites and are given by (from left to right):
$|S\rangle^{\otimes 22}|R^+\rangle|R^+\rangle|R^-\rangle|S\rangle|S\rangle|R^+\rangle|S\rangle^{\otimes 22}$, $|S\rangle^{\otimes 22}|R^-\rangle\,|S\rangle^{\otimes 4}|L^+\rangle|S\rangle^{\otimes 22}$ and $|S\rangle^{\otimes 22}|R^+\rangle\,|S\rangle^{\otimes 4}|L^+\rangle|S\rangle^{\otimes 22}$.

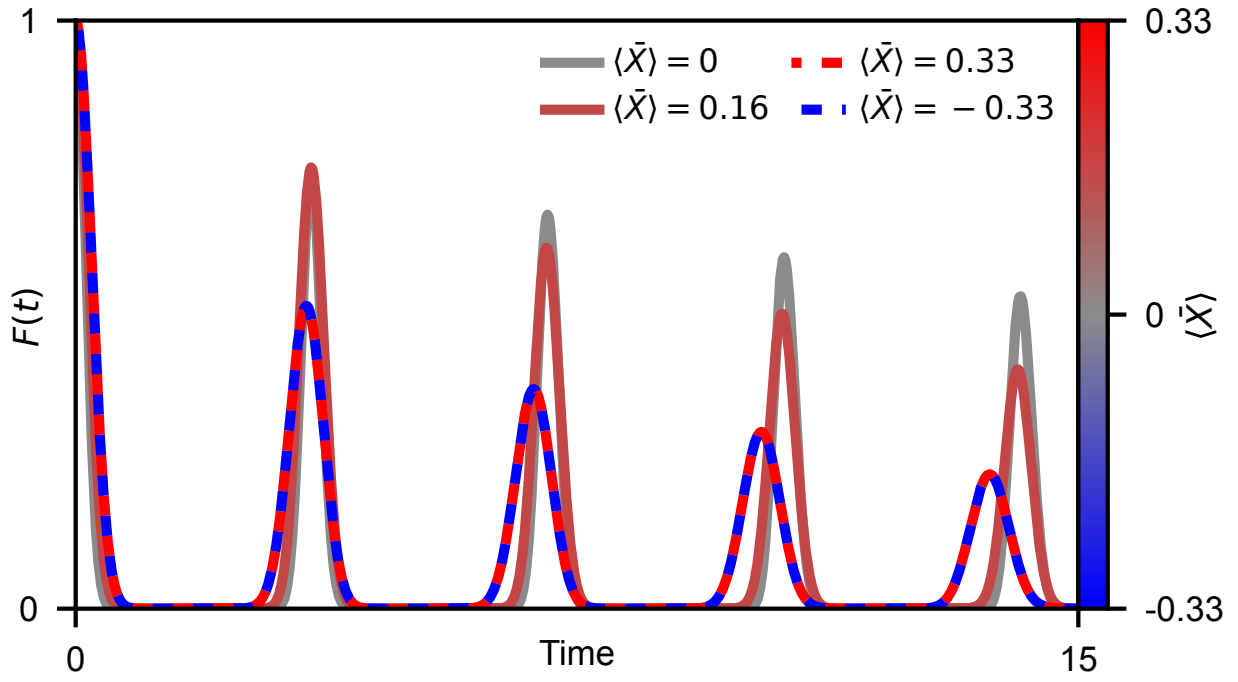

Supplementary Fig. 6. Time-evolution for various scarred states $|\psi\rangle = |R^{\alpha}\rangle^{\otimes 30}$ with different mean energies $\langle \bar{X}\rangle$ (average is taken over all sites $i$) defined on 90 sites. $\langle \bar{X}\rangle = 0$ corresponds to $|S\rangle$ ($\alpha = \pi/2$) unit cells, $\langle \bar{X}\rangle = \pm 0.33$ to $|R^{\pm}\rangle$ ($\alpha = \pm\pi/4$) and $\langle \bar{X}\rangle = 0.16$ to $\alpha = 0.26$. Pronounced periodic fidelity revivals can be observed for scarred states with all energies, while the revivals for scarred states with extremal energy densities are lower. The fidelities are being measured on the subsystem $A = \{31, 32, \ldots 60\}$.

The expansion of an initial state that has a single soliton cell $|R\rangle$ inserted in the scarred background, shown in Supplementary Fig. 7**c**, is, however, more intricate. First, overlaps with this initial state single out a few eigenstates at positive and negative energies. These dominant eigenstates occur at the same energies as the ones responsible for scarring from the $|K\rangle$ state, and several of them are actually the same for both initial states. However, while eigenstates closer to edges of the spectrum have somewhat enhanced overlaps compared to other states at similar energies, visible as "gap" in Supplementary Fig. 7**c** separating them from the rest of the states, this gap is missing near the center of the spectrum. As such, compared to the $|S\rangle$ state, many more eigenstates contribute to the dynamics when a defect cell is introduced. This is expected based on symmetry considerations: for the $|K\rangle$ state, only the $k = 0$ and $k = \pm\frac{2\pi}{3}$ momentum sectors contribute, whereas for single-defect states *all* symmetry sectors are relevant. These results suggest that, while the states with solitons retain high overlap with the eigenstates linked to the scarred background $|S\rangle$, there is no simple subset of eigenstates responsible for soliton dynamics. Although we cannot exclude the possibility that some weak deformation of the PXP model may stabilize solitons and reveal a simpler structure in their eigenstate expansions.

## Supplementary Note 4. TDVP EQUATIONS OF MOTION AND DYNAMICS

The following optimization procedure has been applied in order to find the soliton and scarred angles, denoted by vectorized notations as $\boldsymbol{\theta}^S = (\theta_1^S, \theta_2^S, \theta_3^S)$, and $\boldsymbol{\theta}^R = (\theta_1^R, \theta_2^R, \theta_3^R)$, respectively. The simulation has been performed on a system with a unit cell of size 15, in which the following two sets of angles are defined:

$$\boldsymbol{\theta}^{\text{init}} = (\boldsymbol{\theta}^S, \boldsymbol{\theta}^R, \boldsymbol{\theta}^S, \boldsymbol{\theta}^S, \boldsymbol{\theta}^S), \qquad \boldsymbol{\theta}^{\text{shift}} = (\boldsymbol{\theta}^S, \boldsymbol{\theta}^S, \boldsymbol{\theta}^R, \boldsymbol{\theta}^S, \boldsymbol{\theta}^S).$$

The EOMs for the initial state defined by the angles $\boldsymbol{\theta}^{\text{init}}$ are integrated up to the final time $t = 20$, resulting in a function $\boldsymbol{\theta}(t)$ for $t \in [0, 20]$.

We defined a total loss functional operating on $\boldsymbol{\theta}(t)$, $L$, that contains two parts,

$$L = L_{\text{closed}} + L_{\text{shifted}}.$$

The first part measures how well the trajectory returns to its initial state after the soliton has traveled one full period through the chain, and the second part measures how accurately all angles have shifted by 3 sites to the right after

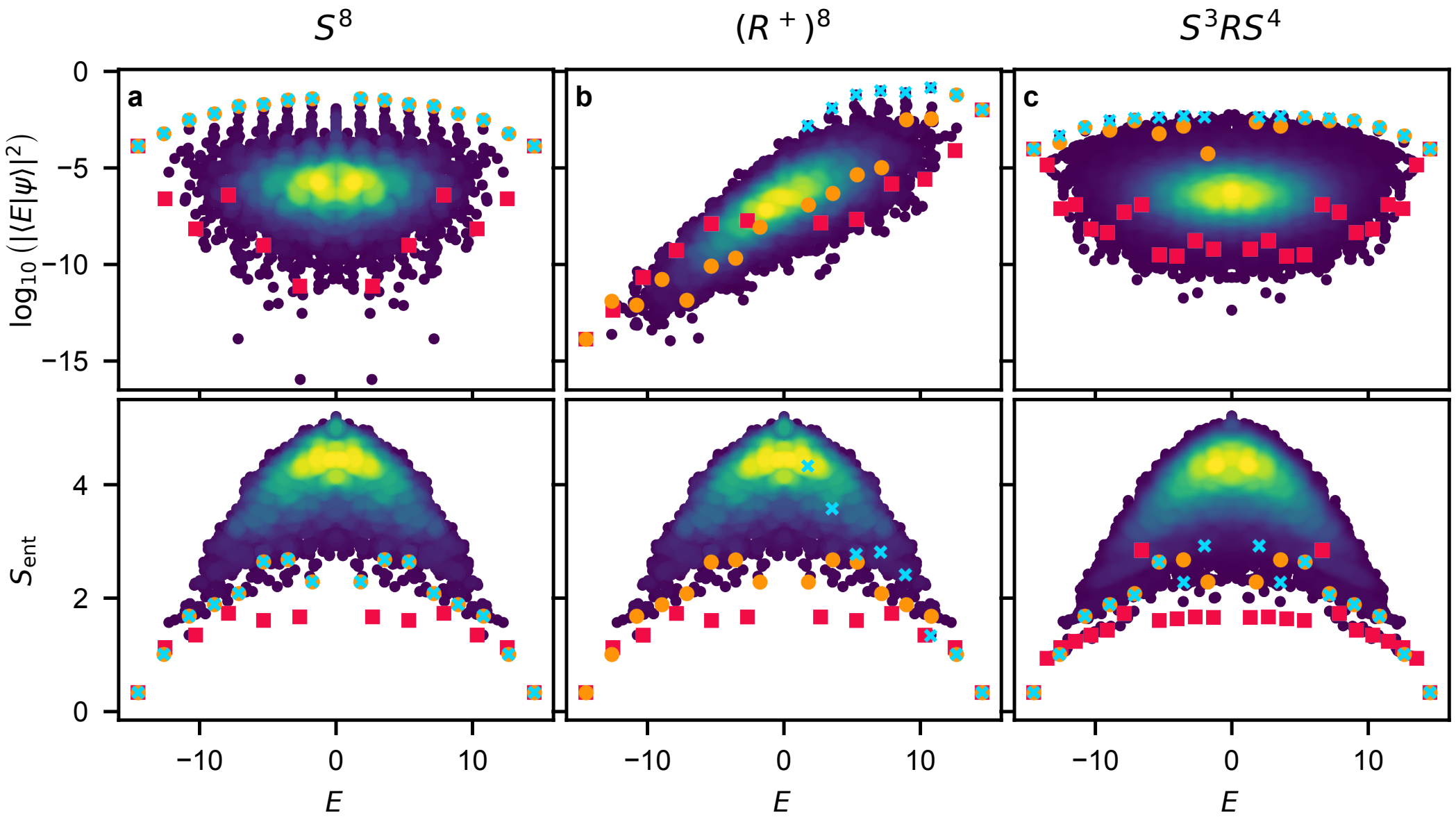

Supplementary Fig. 7. Overlap of eigenstates with different scarred states with and without defects (top row) and half-chain von Neumann entanglement entropy of eigenstates (bottom row). The states are defined on a chain of 24 sites with periodic boundary conditions. Only eigenstates in the symmetry sectors relevant for each state are shown ($k = 0$ and $k = \pm\frac{2\pi}{3}$ in panels **a** and **b** and all momentum sectors for panel **c**). The color indicates the density of data points, going from dark blue to yellow as density increases. The red squares indicate the scarred eigenstates linked to the $|\mathbb{Z}_2\rangle$ state, while the orange circles indicate the ones linked to the entangled $|\mathbb{Z}_3\rangle$ state $|K\rangle = |S\cdots S\rangle$. Finally, the light blue crosses indicate the most likely candidate for scarred eigenstates for the state considered in each panel. They are identified as the eigenstates with the highest overlap in their energy window. For the three different highlighted set of states, the same eigenstates are also highlighted in the same way in the bottom row. **a,** The scarred state $|K\rangle$ has high overlap with a band of approximately evenly spaced eigenstates that is well separated from the bulk, symmetric around zero energy, and has low entanglement. **b,** Scarred state with maximal positive energy with unit cell $|R^+\rangle$ has increased overlaps with higher energy eigenstates. A band of special eigenstates with positive energy shows very high overlap with the scarred states and is well separated from the other eigenstates. **c,** The state $K^{\mathbf{l},\mathbf{r}}$ with a single right-moving defect ($\mathbf{l} = \varnothing$, $\mathbf{r} = \{4\}$) shows several eigenstates with heightened overlap that have low entanglement. However, these states are not very well separated from the bulk of the eigenstates.

one scar revival period. The first part of the loss functional is defined as:

$$L_{\mathrm{closed}} = \min_{t\in[15,20]} \sum_{i=1}^{15} \Delta_i\big(t; \boldsymbol{\theta}^{\mathrm{init}}\big) + P_{\mathrm{closed}}, \tag{7}$$

$$\Delta_i(t; \mathbf{v}) = \left|\cos^2\theta_i(t) - \cos^2 v_i\right| + \left|\sin^2\theta_i(t) - \sin^2 v_i\right|. \tag{8}$$

$$P_{\mathrm{closed}} = \begin{cases} \cos^2\!\big(\frac{\pi}{2}\, d/0.4\big), & d < 0.4, \\ 0, & d \geq 0.4, \end{cases} \tag{9}$$

$$d = \sum_{1\leq i<j\leq 3} \big[\,\big|\,|\cos\theta_i^S| - |\cos\theta_j^S|\,\big| + \big|\,|\sin\theta_i^S| - |\sin\theta_j^S|\,\big|\,\big], \tag{10}$$

where $\Delta_i(t; \mathbf{v})$ is a helper function that defines how we measure the distance between angles. It takes an angle vector $\mathbf{v}$ as input and measures its distance to the time-evolved angles in the $i$-th coordinate. Consequently, the first term of $L_{\mathrm{closed}}$ measures how close the trajectory returns to its initial conditions in the expected time window ($5\cdot 3 \leq t \leq 5\cdot 4$) in which the soliton travels through all 5 cells (time window of scar reviving period was chosen to be $3 \leq t \leq 4$). $P_{\mathrm{closed}}$ is a penalty that is introduced to ensure that the angles of the scarred cells are not all identical to each other, since this leads to solutions that yield relatively low losses but do not show solitonic dynamics. The functional form of $P_{\mathrm{closed}}$ is chosen to ensure a smooth penalty, and the cutoff value 0.4 was empirically found to be a good value.

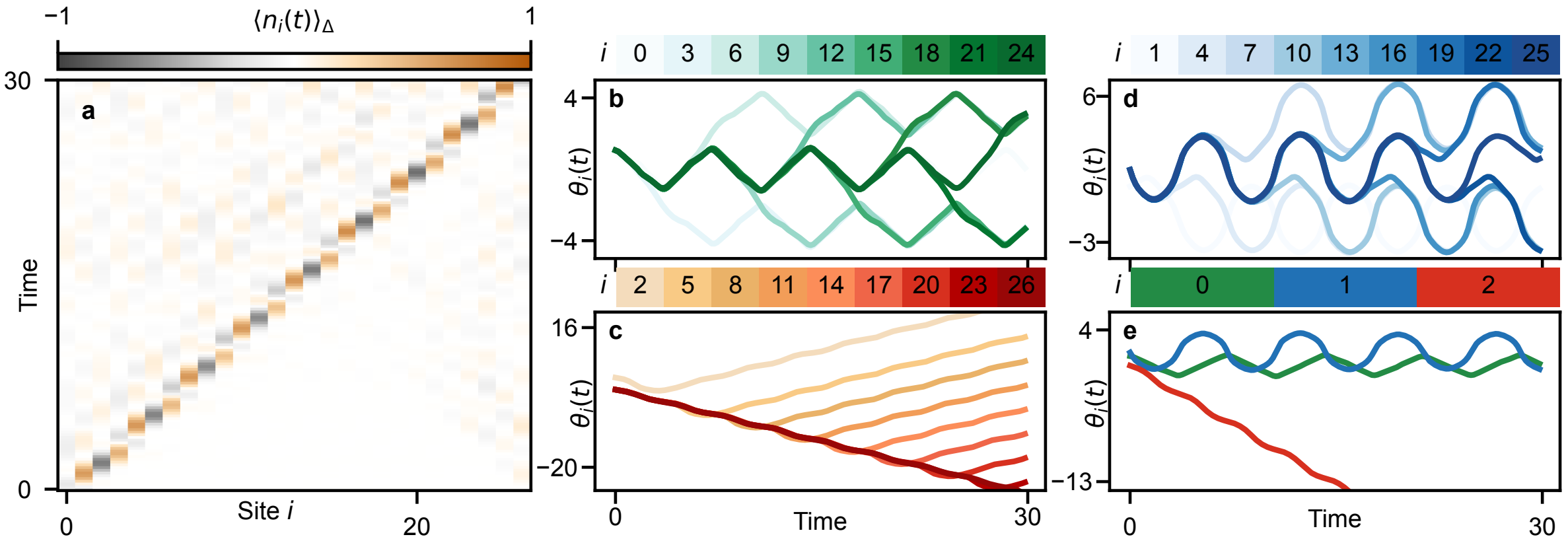

Supplementary Fig. 8. Time evolution of classical equations of motion (9) of an infinite-chain with a 27-site unit cell up to time $t = 30$. The initial state has all sites initialized with $\theta^S_{1,2,3}$ angles except for the sites $\{1, 2, 3\}$ initialized with right-moving soliton $\theta^R_{1,2,3}$ angles. **a,** The characteristic soliton propagation can be observed in the expectation value of the number operator density between states with and without a soliton cell. **b-d** Evolution of angles of the first and second site of each 3-site shows approximately periodic dynamics up to aperiodic shifts induced when the soliton crosses the unit-cell. **e** Periodic dynamics of the three scarred angles for the initial state without the soliton cell.

The second part has a somewhat simpler form as

$$L_{\text{shifted}} = \min_{t\in[3,4]} \sum_{i=1}^{15} \Delta_i\big(t; \boldsymbol{\theta}^{\text{shift}}\big) + P_{\text{shifted}}. \tag{11}$$

$$P_{\text{shifted}} = \begin{cases} 3\cos^2\!\big(\frac{\pi}{2}\, D/0.7\big), & D \leq 0.7, \\ 0, & D > 0.7, \end{cases} \tag{12}$$

$$D = \sum_{a=1}^{3} \big(|\cos^2\theta^S_a - \cos^2\theta^R_a| + |\sin^2\theta^S_a - \sin^2\theta^R_a|\big), \tag{13}$$

where the first term in $L_{\text{shifted}}$ measures the distance between the time-evolved angles and $\boldsymbol{\theta}^{\text{shift}}$ in the time window of the expected scar reviving period. The penalty term $P_{\text{shifted}}$ ensures that the parameters of the soliton cell are different from those of the scarred cells. The function is again chosen to be smooth, and the cutoff value 0.7 was chosen empirically.

We have run this minimization procedure several times and used the parameters that most convincingly produced the solitonic dynamics, which are provided in the Methods section. Numerical integration reveals that these equations feature many singular points, originating from the PXP constraint. Some of these singularities occur close to the angles corresponding to the scarred cells, which makes the optimization very challenging, especially for gradient-based optimizers. We used SciPy's stochastic differential evolution algorithm [6, 7] for this purpose. We note that it is likely that better initial configurations – i.e., initial configurations leading to more coherent solitonic dynamics or even closed trajectories – exist.

In Fig. 5 of the main text, we have established the existence of solitons within our variational description. To get a better understanding of the evolution of the angles $\theta_i(t)$ in a classical dynamical system, we revisit the same initial state (with a smaller unit cell) in Supplementary Fig. 8 and plot the evolution of all angles. In panel **a**, the number operator density clearly shows the coherent propagation of the soliton. Panels **b-c** show the periodic motion of the first and second angles of each unit cell correspondingly, until the soliton passes through, temporarily disturbing this periodic motion, and shifting the angle to a replica of the periodic trajectory. The third angle in each cell shown in panel **d** corresponds to a winding angle whose direction reverses as the soliton passes through. Panel **e** shows the characteristic periodic dynamics of the purely scarred initial state without solitons. As discussed in the main text this state is very close to a point on the trajectory of $|K\rangle$ which is the trajectory discussed in Ref. [8].

In the main text, it has been shown that the classical EOMs (13) and (14) derived via TDVP are capable of capturing solitonic dynamics. The scarred background cell $\theta^S_{1,2,3} = (1.05977133, 1.46820419, 0.0000131027276)$

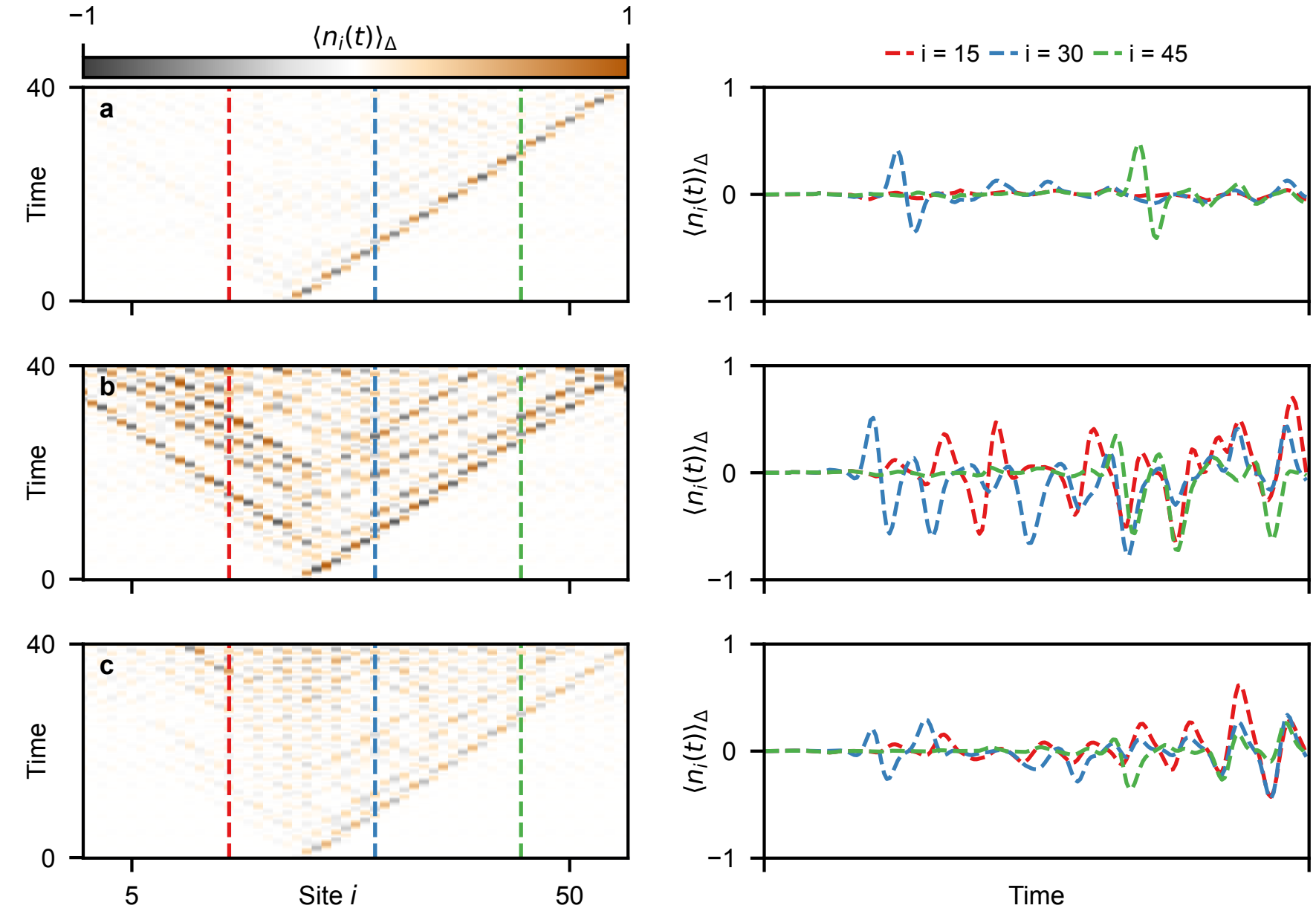

Supplementary Fig. 9. Time evolution of classical EOMs (13) and (14) of the main text for different scarred initial states defined using a 57-site unit cell with a single defect cell inserted at sites $\{22, 23, 24\}$. The left column shows the evolution of the number operator densities with and without the defect cells. The right column shows the evolution of the number operator at three specific sites. **a,** Dynamics of the numerically optimized state shown in the main text with angles of scarred cell given by $\theta^S_{1,2,3} = (1.05977133, 1.46820419, 0.0000131027276)$ and defect cell by $\theta^R_{1,2,3} = (0.990853599, 0.454003269, 3.10125177)$. The defect cell propagates coherently even at later times. **b,** Dynamics of the scarred state $|S\rangle$ corresponding to the angles $\tilde{\theta}^S_{1,2,3} = (0.0, 0.57637521, \pi/2)$ with a right-moving soliton $|R\rangle$ corresponding to $\tilde{\theta}^R_{1,2,3} = (0.0, 2.5069758, 2.64260532)$. Unlike in the fully quantum case, this state shows fast decay of the soliton and subsequent incoherent propagation in both directions. **c,** Dynamics of the scarred state $|S\rangle$ given by $\tilde{\theta}^S_{1,2,3} = (0.0, 0.57637521, \pi/2)$ with right-moving energy-carrying soliton $|R^+\rangle$ given by $\tilde{\theta}^{R^+}_{1,2,3} = (0.0, 2.53577767, 2.22464348)$ and $\tilde{\phi}^{R^+}_{1,2,3} = (0.0, 0.82745493, 2.71076968)$. While coherent propagation is visible at early times, it eventually turns incoherent.

and the right-moving defect cell $\theta^R_{1,2,3} = (0.990853599, 0.454003269, 3.10125177)$ have been numerically optimized to achieve coherent propagation of the solitonic defect. However, the solitonic defects $|L, R\rangle$ and $|L^\pm, R^\pm\rangle$ embedded in the scarred background state $|K\rangle$ can also be faithfully represented via the Ansatz in Eq. (11) of the main text, allowing their time-evolution to be studied within the semiclassical description. The angles for the scarred cell $|S\rangle$ are given by $\tilde{\theta}^S_{1,2,3} = (0.0, 0.57637521, \pi/2)$, for $|L\rangle$ and $|R\rangle$ by $\tilde{\theta}^L_{1,2,3} = (0.0, 2.50697563, 0.49898734)$ and $\tilde{\theta}^R_{1,2,3} = (0.0, 2.5069758, 2.64260532)$. The energy-carrying soliton cells, which have phase angles $\phi \neq \pi/2$, are given by $\tilde{\theta}^{R^+}_{1,2,3} = (0.0, 2.53577767, 2.22464348)$, $\tilde{\phi}^{R^+}_{1,2,3} = (0.0, 0.82745493, 2.71076968)$ and $\tilde{\theta}^{R^-}_{1,2,3} = (0.0, 2.53577767, 2.22464348)$, $\tilde{\phi}^{R^-}_{1,2,3} = (0.0, 2.31413787, 0.43082226)$ as well as $\tilde{\theta}^{L^+}_{1,2,3} = (0.0, 2.53577818, 0.91695008)$, $\tilde{\phi}^{L^+}_{1,2,3} = (\pi/2, 2.31414114, 0.43082552)$ and $\tilde{\theta}^{L^-}_{1,2,3} = (0.0, 2.53577773, 0.91694925)$, $\tilde{\phi}^{L^-}_{1,2,3} = (0.0, 0.82745289, 2.71076687)$.

In Supplementary Fig. 9**b** we show that $|R\rangle$ embedded in the scarred state background cells $|S\rangle$ initially shows solitonic dynamics but quickly loses coherence compared to the numerically optimized state in Supplementary Fig. 9**a**. The energy-carrying soliton (Supplementary Fig. 9**c**) shows coherent propagation for longer times but still does not match the coherent propagation of the numerically optimized state.

A numerically optimized left-moving defect cell is given by the angles $\theta^L_{1,2,3} = (4.13244625, 3.59559592, 6.24284442)$. Numerically optimized left- and right-moving solitons interact strongly upon collision, deform the scarred background, and lose their coherence as shown in Supplementary Fig. 10**a**. The collision between the energy-carrying solitons $|L^+\rangle$ and $|R^+\rangle$ appears to be more stable (see Supplementary Fig. 10**b**), as solitons continue to propagate without a phase shift and no significant decrease in energy after the collision. Nevertheless, the collision strongly distorts the scarred background.

So far, we have considered states with solitonic excitations that show coherent propagation under the exact dynamics of the quantum model, and have investigated their evolution in the system of classical EOMs. Conversely, we will now take the soliton that we have numerically optimized in the semiclassical limit and investigate its dynamics under unitary time-evolution of the quantum model. Using the MPS Ansatz given in Eq. (11) of the main text (fixing all phase angles $\phi = \pi/2$), we define the scarred state via the numerically optimized angles $\theta^S_{1,2,3}$ as

$$|\psi_S\rangle = \sum_{\{\sigma\}} V_L \prod_{l=0}^{N/3-1} A^{\sigma_{1+3l}}(\theta_1^{\mathrm{S}})\, A^{\sigma_{2+3l}}(\theta_2^{\mathrm{S}})\, A^{\sigma_{3+3l}}(\theta_3^{\mathrm{S}})\, V_R\, |\sigma_1\sigma_2\ldots\sigma_N\rangle\,, \tag{14}$$

and the state with the incorporated soliton defect cell $\theta^R_{1,2,3}$ as

$$\begin{aligned} |\psi_R\rangle = \sum_{\{\sigma\}} & V_L \Big[\prod_{l=0}^{l_0-1} A^{\sigma_{1+3l}}(\theta_1^{\mathrm{S}})\, A^{\sigma_{2+3l}}(\theta_2^{\mathrm{S}})\, A^{\sigma_{3+3l}}(\theta_3^{\mathrm{S}})\Big] \\ & \times A^{\sigma_{3l_0+1}}(\theta_1^{\mathrm{R}})\, A^{\sigma_{3l_0+2}}(\theta_2^{\mathrm{R}})\, A^{\sigma_{3l_0+3}}(\theta_3^{\mathrm{R}}) \\ & \times \Big[\prod_{l=l_0+1}^{N/3-1} A^{\sigma_{1+3l}}(\theta_1^{\mathrm{S}})\, A^{\sigma_{2+3l}}(\theta_2^{\mathrm{S}})\, A^{\sigma_{3+3l}}(\theta_3^{\mathrm{S}})\Big] V_R\, |\sigma_1\sigma_2\ldots\sigma_N\rangle\,, \end{aligned} \tag{15}$$

where the physical indices $\sigma$ label the on-site Hilbert space basis $(\uparrow, \downarrow)$. $V_L$ and $V_R$ are the left and right boundary vectors, chosen to be $(1,0)^T$ and $(1,1)$ respectively, and $l_0$ defines the position of the soliton cell. The cells of the MPSs are weakly entangled and have no product-like structure. We measure their similarity with regard to the cells that show scarred and solitonic under unitary time-evolution of the quantum model by measuring the projector expectation values. It turns out the numerically optimized scarred state is locally almost identical to the state $|\bar{S}\rangle$ that occurs along the scarred trajectory. Indeed, we find that in a system with 153 sites, $\langle\psi_S|P_{\bar{S},75}|\psi_S\rangle \approx 0.99$, where $P_{\bar{S},j} = |\bar{S}\rangle\langle\bar{S}|$ is the projector on the $\bar{S}$ cell at position $j$ to $j+2$. Analogously, we are measuring the expectation value of $P_{R,75} = |R\rangle\langle R|$ with respect to $|\psi_R\rangle$. The position of the defect angles $\{3l_0+1, 3l_0+2, 3l_0+3\}$ coincides with $\{75, 76, 77\}$ and the expectation value yields $\langle\psi_R|P_{R,75}|\psi_R\rangle \approx 0.56$. Finally, time-evolving $|\psi_R\rangle$ under unitary dynamics of the quantum model leads the defect to propagate coherently to the right with only modest deformation of the scarred background (Supplementary Fig. 11). Large peaks of the translation fidelities reveal that the defect cell approximately maintains its structure. The revivals of all $F_m(t)$ at multiples of $T$ reveal finite overlap of the defect and the scarred cell. Solitonic excitations in the classical EOMs appear to be more stable when they have finite overlap with the scarred cell.

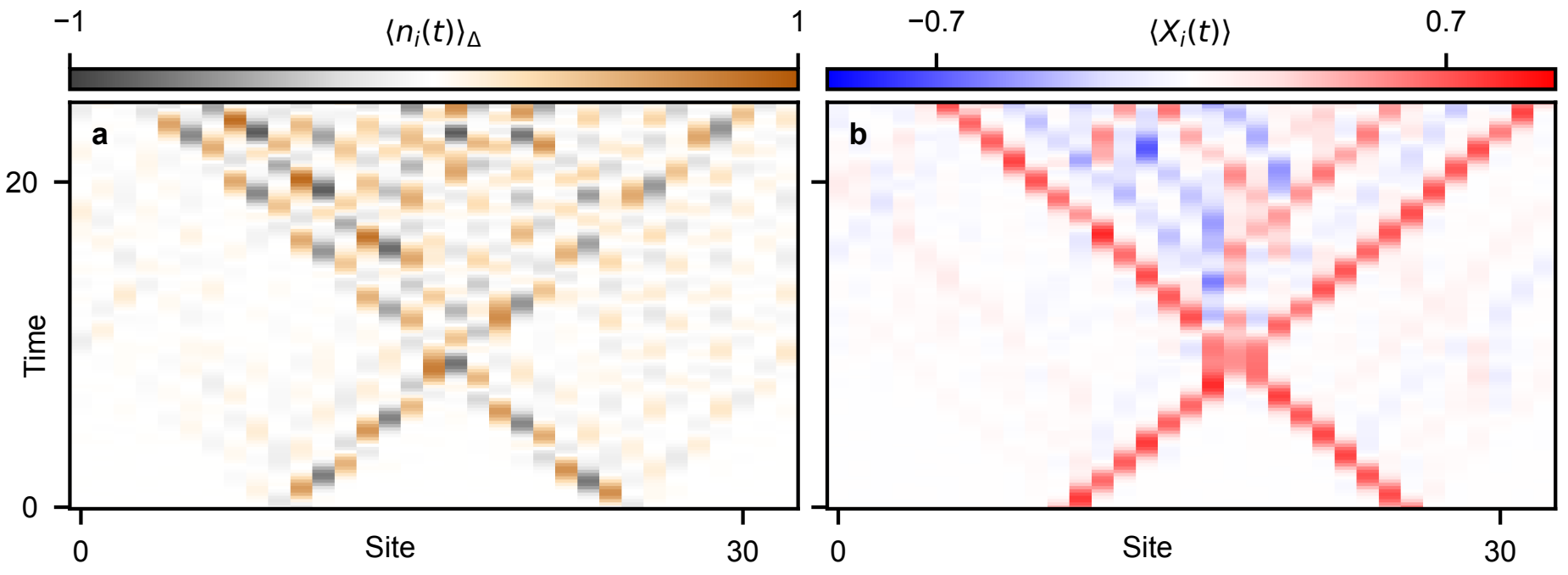

Supplementary Fig. 10. Time evolution of classical EOMs (13), (14) of the main text showing collisions of solitons propagating in opposite directions on top of a scarred background state defined on 33 sites. **a,** Numerically optimized zero-energy solitons with the scarred background cells defined via the angles $\theta^S_{1,2,3} = (1.05977133, 1.46820419, 0.0000131027276)$ and a right-moving soliton cell defined via the angles $\theta^R_{1,2,3} = (0.990853599, 0.454003269, 3.10125177)$ on the sites $\{10, 11, 12\}$ and a left-moving soliton with angles $\theta^L_{1,2,3} = (4.13244625, 3.59559592, 6.24284442)$ on the sites $\{25, 26, 27\}$. Solitons are strongly interacting during the collision and perturb the scarred background. **b,** Energy solitons on top of the usual scarred background cells $|S\rangle$ with angles $\tilde{\theta}^S_{1,2,3} = (0.0, 0.57637521, \pi/2)$ and a right-moving energy soliton cell $|R^+\rangle$ given via $\tilde{\theta}^{R^+}_{1,2,3} = (0.0, 2.53577767, 2.22464348)$ and $\tilde{\phi}^{R^+}_{1,2,3} = (0.0, 0.82745493, 2.71076968)$, positioned at sites $\{10, 11, 12\}$ and a left-moving energy soliton cell $|L^+\rangle$ given via $\tilde{\theta}^{L^+}_{1,2,3} = (0.0, 2.53577818, 0.91695008)$ and $\tilde{\phi}^{L^+}_{1,2,3} = (\pi/2, 2.31414114, 0.43082552)$, positioned at sites $\{25, 26, 27\}$.

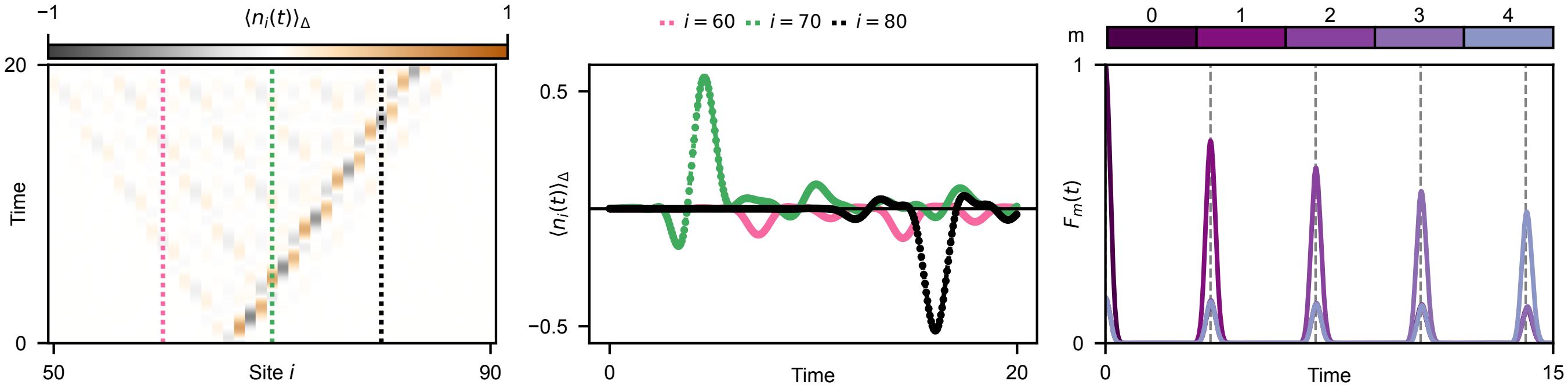

Supplementary Fig. 11. Dynamics of numerically optimized soliton state in the full quantum model defined on 150 sites. Evolving the MPS given by the scarred background cells with the angles $\theta^S_{1,2,3} = (1.05977133, 1.46820419, 0.0000131027276)$ and the soliton angles given by $\theta^R_{1,2,3} = (0.990853599, 0.454003269, 3.10125177)$ and positioned on $\{82, 83, 84\}$ leads to coherent propagation of the soliton. This manifests in the difference between the number operator of the state with and without a defect cell, as well as high translation fidelity peaks occurring at multiples of the revival period $T \approx 3.52$.

In Ref. [9], TDVP equations for the PXP model with an arbitrary unit cell size are also derived. For spin-1/2, the Ansatz used in that work is identical to ours, except for a factor two in the definition of the $\theta_j$ angle and a minus sign for the $\phi_j$ angles.

Using the results of that work with $\Delta_i = 0$ and $\Omega_i = 1$ for all $i$, $J = 1/2$, one gets the following results:

$$\partial_t\theta_i = 2\frac{\tan(\phi_i)}{\sin(\theta_i)}\tilde{h}_{s^x_i} + \frac{\eta_{i-1}\tan(\phi_{i-1})\tan\left(\frac{\theta_i}{2}\right)}{\eta_i}\tilde{h}_{s^x_{i-1}} \tag{16}$$

$$\partial_t\phi_i = \frac{\tilde{h}_{s^x_i}}{\sin^2\left(\frac{\theta_i}{2}\right)} - 2\frac{\tan\left(\frac{\theta_{i+1}}{2}\right)}{\eta_{i+1}}\tilde{R}_{\theta_{i+1}} - \frac{4}{\eta_i\sin(\theta_i)}\tilde{R}_{\theta_i}\ , \tag{17}$$

where

$$\tilde{h}_{s^x_i} = \sin(\theta_i)\cos(\phi_i)\cos\left(\frac{\theta_{i+1}}{2}\right) \tag{18}$$

$$\tilde{R}_{\theta_i} \equiv \frac{1}{4}\eta_{i-1}\sin\left(\frac{\theta_i}{2}\right)\sin(\theta_{i-1})\cos(\phi_{i-1}) + \frac{\eta_i}{2}\cos(\phi_i)\cos\left(\frac{\theta_{i+1}}{2}\right)\ . \tag{19}$$

As noted in Ref. [9] below Eq. (20), for spin-1/2 all $\tilde{c}_i = 0$.

The $\eta_i$ in these equations can be linked to the $\langle n_i\rangle$ we define in Eq. (15) of the main text as $\langle n_i\rangle = \sin^2\left(\frac{\theta_i}{2}\right)\eta_i$. This can be seen clearly by rewriting the $\eta_i$ given for small values of $L$ in Eq. (10) of Ref. [9] in a familiar form as

$$\eta_i = \begin{cases} \frac{1}{1-\sin^2\left(\frac{\theta_i}{2}\right)} & L = 1 \\ \frac{1-\sin^2\left(\frac{\theta_{i-1}}{2}\right)}{1-\left[-\sin^2\left(\frac{\theta_{i-1}}{2}\right)\right]\left[-\sin^2\left(\frac{\theta_i}{2}\right)\right]} & L = 2 \\ \frac{1-\sin^2\left(\frac{\theta_{i-2}}{2}\right)\left[1-\sin^2\left(\frac{\theta_{i-1}}{2}\right)\right]}{1-\left[-\sin^2\left(\frac{\theta_{i-2}}{2}\right)\right]\left[-\sin^2\left(\frac{\theta_{i-1}}{2}\right)\right]\left[-\sin^2\left(\frac{\theta_i}{2}\right)\right]} & L = 3\ . \end{cases} \tag{20}$$

Substituting the full expressions for $\eta_i$ and $\tilde{h}_{s^x_i}$, we can further simplify Eq. (16) as

$$\frac{\partial_t\theta_i}{2} = \sin(\phi_i)\cos\left(\frac{\theta_{i+1}}{2}\right) + \frac{\langle n_{i-1}\rangle}{\langle n_i\rangle}\sin(\phi_{i-1})\sin\left(\frac{\theta_{i-1}}{2}\right)\cot\left(\frac{\theta_{i-1}}{2}\right)\sin^3\left(\frac{\theta_i}{2}\right)\ . \tag{21}$$

This matches exactly with our independently derived Eq. (13) of the main text up to the transformations $\theta_i/2 \to \theta_i$ and $\phi_i \to -\phi_i$ which stems from the difference in the Ansätze used. We can perform the exact same procedure on

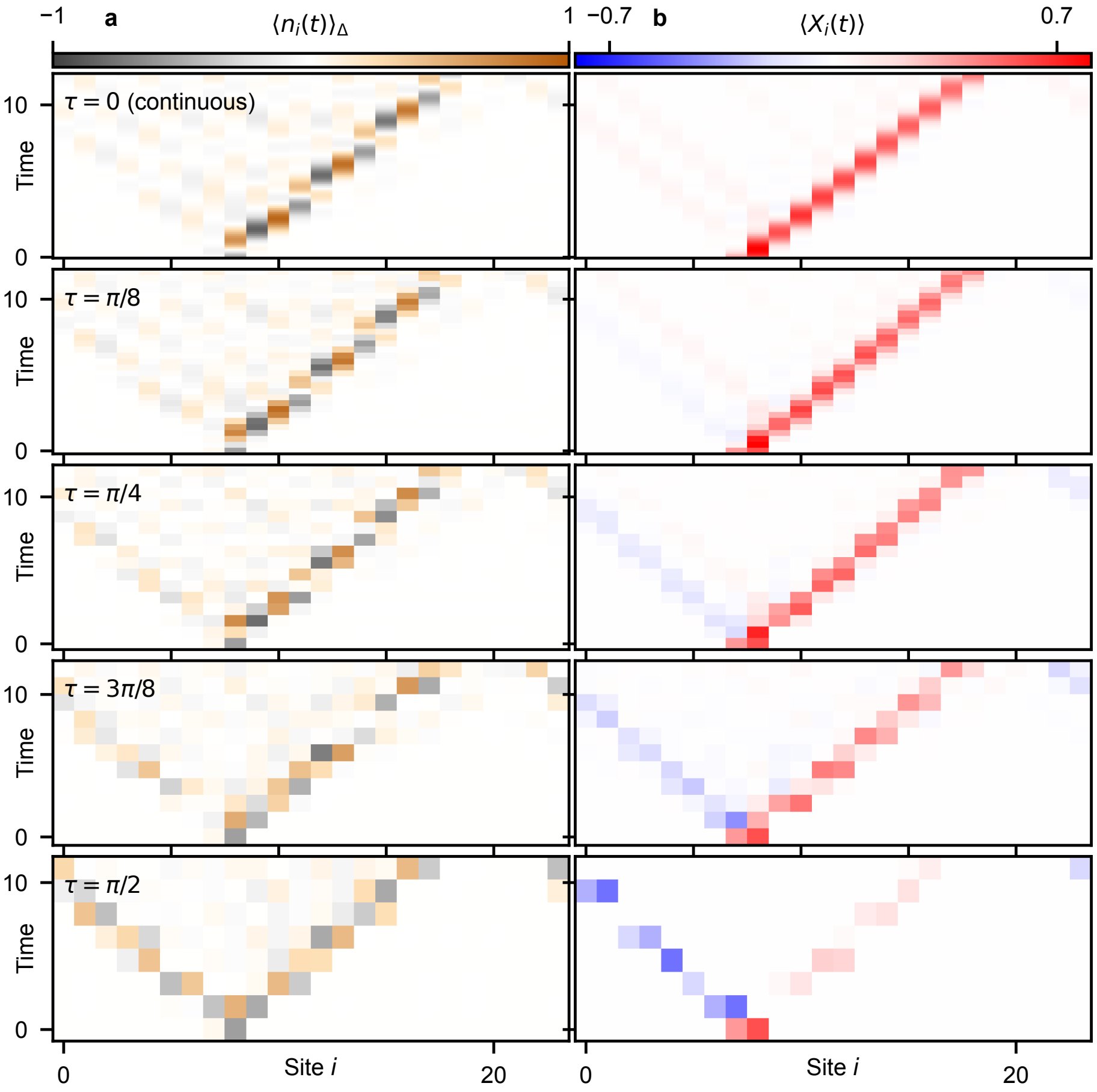

Supplementary Fig. 12. Time evolution of solitons under the Trotterized PXP Hamiltonian given in Eq. (23) with step sizes interpolated between the time-continuous PXP model ($\tau = 0$) up to the integrable point ($\tau = \pi/2$) **a,** The number operator density difference of the initial $|S\rangle^{\otimes 2}|R\rangle|S\rangle^{\otimes 5}$ state reveals coherent propagation for evolution with all considered Trotter step sizes, but the defect loses its chirality upon increasing $\tau$ and spreads nearly symmetrically in both directions for $\tau = \pi/2$. **b,** The energy-soliton state $|S\rangle^{\otimes 2}|R^{+}\rangle|S\rangle^{\otimes 5}$ also shows an increasingly symmetric propagation of $\langle X_i(t)\rangle$ (with opposite signs) with increasing Trotter step size.

Eq. (17). While this requires substantially more steps, it is straightforward and leads to

$$\begin{aligned}-\partial_t\phi_i = \cos(\phi_i)\left[\sin^3\left(\frac{\theta_{i+1}}{2}\right)\frac{\langle n_i\rangle\cot\left(\frac{\theta_i}{2}\right)}{\langle n_{i+1}\rangle\cot\left(\frac{\theta_{i+1}}{2}\right)} - 2\cos\left(\frac{\theta_{i+1}}{2}\right)\cot(\theta_i)\right] \\ + \cos\left(\frac{\theta_{i+2}}{2}\right)\tan\left(\frac{\theta_{i+1}}{2}\right)\cos(\phi_{i+1}) + \sin\left(\frac{\theta_i}{2}\right)\frac{\langle n_{i-1}\rangle\cot\left(\frac{\theta_{i-1}}{2}\right)}{\langle n_i\rangle\cot\left(\frac{\theta_i}{2}\right)}\cos(\phi_{i-1}),\end{aligned} \tag{22}$$

which now matches with Eq. (15) of the main text up to the same transformations $\theta_i/2 \to \theta_i$ and $\phi_i \to -\phi_i$.

## Supplementary Note 5. FLOQUET DYNAMICS

Recent works have investigated Floquet realizations of the Rydberg atom chain, with special attention to classical cellular automaton (CA) limits, which revealed surprising connections to quantum many-body dynamics [10–12]. Especially interesting to us in this context is Ref. [10], where the authors analyzed the RCA201/Floquet-PXP model, which is an integrable classical CA. Interestingly, this model exhibits solitons that show ballistic transport and interact non-trivially upon collisions with a phase shift but otherwise re-emerge intact after the collision. They can be understood as domain walls between three vacuum states. These solitons in the classical model bear resemblance in certain aspects to the quasi-solitons that we study in the time-continuous PXP model.

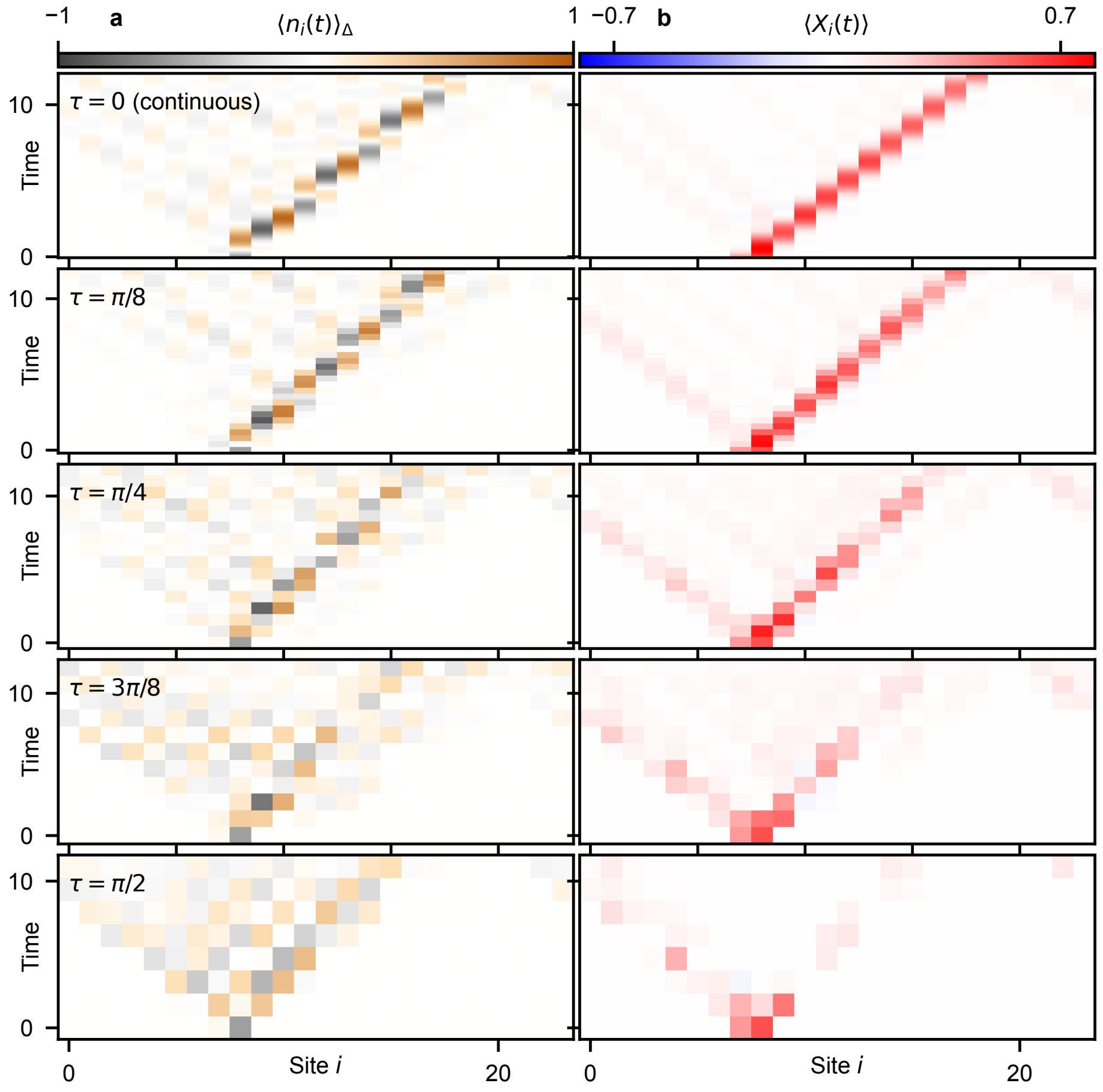

Supplementary Fig. 13. Dynamics of solitons under evolution of the Trotterized PXP Hamiltonian given in Eq. (24) with step sizes interpolated between the time-continuous PXP model ($\tau = 0$) up to the integrable point ($\tau = \pi/2$) **a,** The number operator density difference of the initial $|S\rangle^{\otimes 2}|R\rangle|S\rangle^{\otimes 5}$ shows less coherent propagation for larger step sizes but still appears to move predominantly to right up to $\tau = \pi/2$. **b,** The soliton dynamics shown in form of the expectation value $\langle X_i(t)\rangle$ of the state $|S\rangle^{\otimes 2}|R^+\rangle|S\rangle^{\otimes 5}$ becomes increasingly incoherent for larger Trotter step sizes.

The CA point can be obtained by a Trotter discretization of the time-continuous PXP dynamics. Concretely, one splits the PXP Hamiltonian $H = \sum_j P_{j-1}X_jP_{j+1}$ into even- and odd-site parts, $H = H_e + H_o$, with $H_{e(o)} = \sum_{j\ \mathrm{even(odd)}} P_{j-1}X_jP_{j+1}$, and defines the Floquet operator:

$$U_\tau \;=\; U^e_\tau U^o_\tau \;=\; e^{-iH_e\tau}\, e^{-iH_o\tau}, \tag{23}$$

where $\tau$ is the Trotter step size. For $\tau \to 0$ one recovers the time-continuous evolution, while $\tau = \pi/2$ yields the integrable classical CA that maps classical states to classical states. A priori, there is no reason why the soliton dynamics of the CA should leave an imprint in the time-continuous PXP model, since $\tau = \pi/2$ is far beyond the perturbative regime. However, in Ref. [12] the authors found surprising connections between the quantum many-body scars associated with the $|\mathbb{Z}_2\rangle$ state and some eigenstates of the PXP CA. This connection was established by treating $\tau$ as an interpolation parameter and demonstrating an adiabatic connection between eigenstates of the time-continuous PXP model and eigenstates of the CA, at least for QMBSs close to zero energy. Furthermore, Ref. [12] studied initial product states that host solitons under the CA time evolution and tracked their dynamics as a function of $\tau$, finding that certain signatures of the soliton dynamics persist down to small Trotter steps.

One can ask how the quasi-solitons found in our work change when interpolating between the time-continuous PXP model and the cellular automaton point by varying the Trotter step size. However, the initial states at play here are not product states but weakly entangled states, which complicates matters. To test whether the dynamics of these soliton states in the time-continuous model are connected to the solitons in the CA, one would, in principle, have to interpolate the Trotter step while at the same time deforming the entangled soliton state into a CA soliton product state. It is not clear how this should be done: this is not a simple one-dimensional interpolation, but rather an arbitrary trajectory in the exponentially large Hilbert space that starts at $\tau = 0$ from the soliton state in the time-continuous PXP model and ends at $\tau = \pi/2$ at a CA soliton product state. As already mentioned, there is no physical argument for why these should be related, since the change in the Trotter step size is far beyond the perturbative regime. Potential explorations in this direction are beyond the scope of this work and are left for future research. Here we only take a first step and explore how the dynamics of our states with a single soliton $|R\rangle$ and $|R^+\rangle$ embedded in the scarred background change when interpolating from the time-continuous limit $\tau = 0$ to the integrable point at $\tau = \pi/2$ while leaving the state invariant. This exploration is also relevant for potential implementations on a quantum computer. In Supplementary Fig. 12 we see that the chiral dynamics of the soliton persists up to large Trotter steps ($\tau = \pi/4$); while the dynamics remains coherent up to $\tau = \pi/2$, it becomes increasingly less chiral, and the defect propagates in both directions.

It is also not clear whether the Trotterization performed in [12] is the best choice for uncovering a connection between our solitons in the time-continuous model and the cellular automaton. In that work, this Trotterization was used to establish a connection between the scarring of the $|\mathbb{Z}_2\rangle$ state and the CA. However, the Trotterization is not unique, and it could be that other decompositions of the Hamiltonian establish a stronger connection to our solitons in the time-continuous PXP model. Another possible way of decomposing the Hamiltonian would be $H = H_1 + H_2 + H_3$ with $H_r = \sum_{j \bmod 3 = r-1} P_{j-1} X_j P_{j+1}$ $r \in \{1, 2, 3\}$, leading to the Floquet operator:

$$U_\tau^{(3)} = e^{-iH_3\tau}\, e^{-iH_2\tau}\, e^{-iH_1\tau}, \tag{24}$$

for a time step $\tau$. Likewise, this becomes a cellular automaton at $\tau = \pi/2$. Starting from the fully polarized down state $|0\rangle = |{\downarrow\downarrow\downarrow}\cdots\rangle$, the evolution exhibits a simple period-4 orbit that cycles through translated $|\mathbb{Z}_3\rangle = |{\downarrow\downarrow\uparrow\downarrow\downarrow\uparrow}\ldots\rangle$ configurations $|0\rangle \xrightarrow{U^{(3)}_{\pi/2}} T\,|\mathbb{Z}_3\rangle \xrightarrow{U^{(3)}_{\pi/2}} T^2\,|\mathbb{Z}_3\rangle \xrightarrow{U^{(3)}_{\pi/2}} |\mathbb{Z}_3\rangle \xrightarrow{U^{(3)}_{\pi/2}} |0\rangle$. However, we have not explored potential quasi-particle dynamics in this CA version. Instead, we again compare a simple interpolation in Trotter step size between the time-continuous PXP model and the $\tau = \pi/2$ point, analogous to the even/odd Trotterization above, for the solitons $|R\rangle$ and $|R^+\rangle$ in Supplementary Fig. 13. It seems that the defect propagation in both cases becomes less coherent for increasing step sizes $\tau$ while $|R\rangle$ remains somewhat chiral up to $\tau = \pi/2$.

In summary, the investigation of two different Trotterizations shows that the even/odd scheme has coherently propagating, however non-chiral defects related to our soliton initial state, whereas the period-3 Trotterization leads to decoherence of solitons. We note that a more thorough investigation would be required to establish any potential meaningful connections between our solitons in the time-continuous PXP model and solitons in Floquet realizations of the PXP model.

## Supplementary Note 6. REALIZATION OF SOLITONS IN RYDBERG ATOM ARRAYS

All results so far have been obtained for the idealized PXP Hamiltonian in Eq. (1) of the main text, which approximates the Rydberg chain in the regime of strong nearest-neighbor interactions and weak residual interactions beyond. Under realistic conditions, however, the Rydberg blockade is not perfect, and van der Waals interactions extend beyond nearest neighbors. To examine the stability of the solitons in a more realistic scenario, we simulate their dynamics under the full two-level Rydberg Hamiltonian of the form:

$$\frac{H}{\hbar} = \frac{\Omega}{2}\sum_i X_i - \Delta \sum_i n_i + \sum_{i<j} V_{ij}\, n_i n_j, \qquad V_{ij} = \frac{V_1}{|i-j|^6}. \tag{25}$$

Physical parameters of this Hamiltonian correspond to the Rabi frequency $\Omega$, detuning strength $\Delta$, and repulsive van der Waals interactions $V_{ij}$ between two Rydberg excitations. Our numerical simulations below consider parameters $\Omega = 2 \times 2\pi$ MHz, $V_1 = 14 \times 2\pi$ MHz $= 7\Omega$, and $\Delta = 0.017 \times 2\pi$ MHz $= 0.0085\Omega$. As proposed in Ref. [13], the non-zero value of the detuning $\Delta$ serves to mitigate the effects of long-range interactions.

We compare the time evolution generated by the PXP Hamiltonian and by Eq. (25) for a single right-moving soliton $|R\rangle$, as well as for an energy-carrying soliton $|R^+\rangle$ in terms of local experimentally observable quantities such as the number operator density and kinetic energy density defined as $\langle X_i^k(t)\rangle = \langle\psi(t)|X_i|\psi(t)\rangle$, as well as the translation fidelity to get a better understanding of their decay. Note that we omit projectors in the definition of the kinetic energy density since the Rydberg blockade can be violated under time evolution generated by Eq. (25). Moreover, note that $\sum_i X_i^k$ is not a conserved quantity for the full Rydberg Hamiltonian.

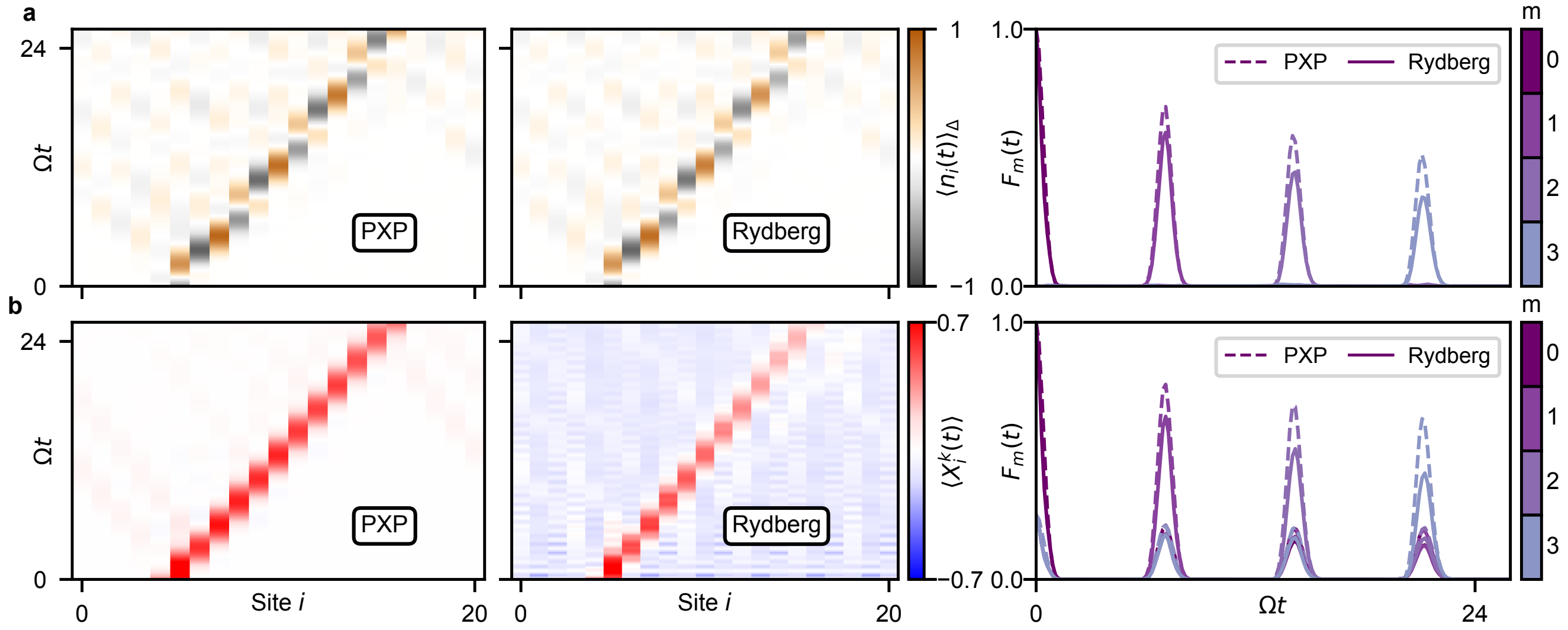

Supplementary Fig. 14. Comparison of unitary time-evolution of solitonic states under the full Rydberg Hamiltonian given in Eq. (25) and the PXP Hamiltonian given in Eq. (1) of the main text performed with sparse Krylov methods. **a,** The number operator density difference of the initial state $|S\rangle|R\rangle\,|S\rangle^{\otimes 5}$ evolved under the PXP Hamiltonian (left) and the full Rydberg Hamiltonian (middle) shows in both cases coherent propagation of the solitonic defect. The translation fidelity revivals (right) for the full Rydberg Hamiltonian are slightly smaller than for the PXP model. **b,** The kinetic energy density of the initial state $|S\rangle|R^+\rangle\,|S\rangle^{\otimes 5}$ evolved under the PXP Hamiltonian (left) and the full Rydberg Hamiltonian (middle) show coherent dynamics of the solitonic defect with a somewhat faster decay under evolution of the Rydberg Hamiltonian. The translation fidelity revivals (right) for the full Rydberg Hamiltonian are also slightly smaller.

In Supplementary Fig. 14, it can be seen that the number operator density between the PXP Hamiltonian and the full Rydberg Hamiltonian seems visually almost identical. The energy density $\langle X_i(t)\rangle$ shows a slightly faster decay under time evolution of the full Rydberg Hamiltonian. Furthermore, we can see that even the translation fidelities for the solitons under the time evolution with the full Rydberg Hamiltonian are comparable to those of the PXP model. These results demonstrate that although Rydberg arrays only approximately realize the PXP model, the dynamics of solitons is only mildly affected by these imperfections. Therefore, the solitons presented in our work can be implemented and studied experimentally in Rydberg atom arrays.

## Supplementary Note 7. SOLITONS IN THE PPXPP MODEL

It was shown that generalizations of the PXP model with larger blockade radii host scarred states as well [2]. Here, we focus on the specific generalization of PXP to blockade of radius two, denoted as the PPXPP model:

$$H = \sum_{i=1}^{N} P_{i-2}P_{i-1}X_iP_{i+1}P_{i+2}. \tag{26}$$

This model requires $|{\uparrow}\rangle$ spin states to be separated by at least two sites, and can be approximately realized by a Rydberg atom array in a triangular ladder geometry [2]. In our previous work, we demonstrated that this model hosts at least two scarred states with unit cell sizes of 4, 5, and 6. Notably, we do not find the soliton excitations on top of scarred state with unit cell sizes 4 and 6, suggesting that the size of the unit cell has to be *odd* as a prerequisite for soliton existence.

For the 5-sites unit cell, we consider the following scarred initial state:

$$|K_5\rangle = \bigotimes_{k=1}^{N/5} |S_5\rangle\,, \qquad |S_5\rangle = \frac{1}{\sqrt{\beta^2+2}}|{\downarrow\downarrow}\rangle(\beta|{\uparrow\downarrow\downarrow}\rangle + |{\downarrow\uparrow\downarrow}\rangle + |{\downarrow\downarrow\uparrow}\rangle), \tag{27}$$

and search for defect cells of 5 sites that would feature coherent right-propagation when initialized in the background of $|S_5\rangle$ cells. In contrast to the PXP solitons, which can be approximately understood as translated unit cells of scarred initial state at quarter and three-quarter periods, a similar intuition does not work for the PPXPP model. Indeed, we observe that appropriately shifted unit cells from the state $|K_5(T/4)\rangle$ and $|K_5(3T/4)\rangle$ violate the blockade condition and cannot be inserted into the background of $|S_5\rangle$ cells.

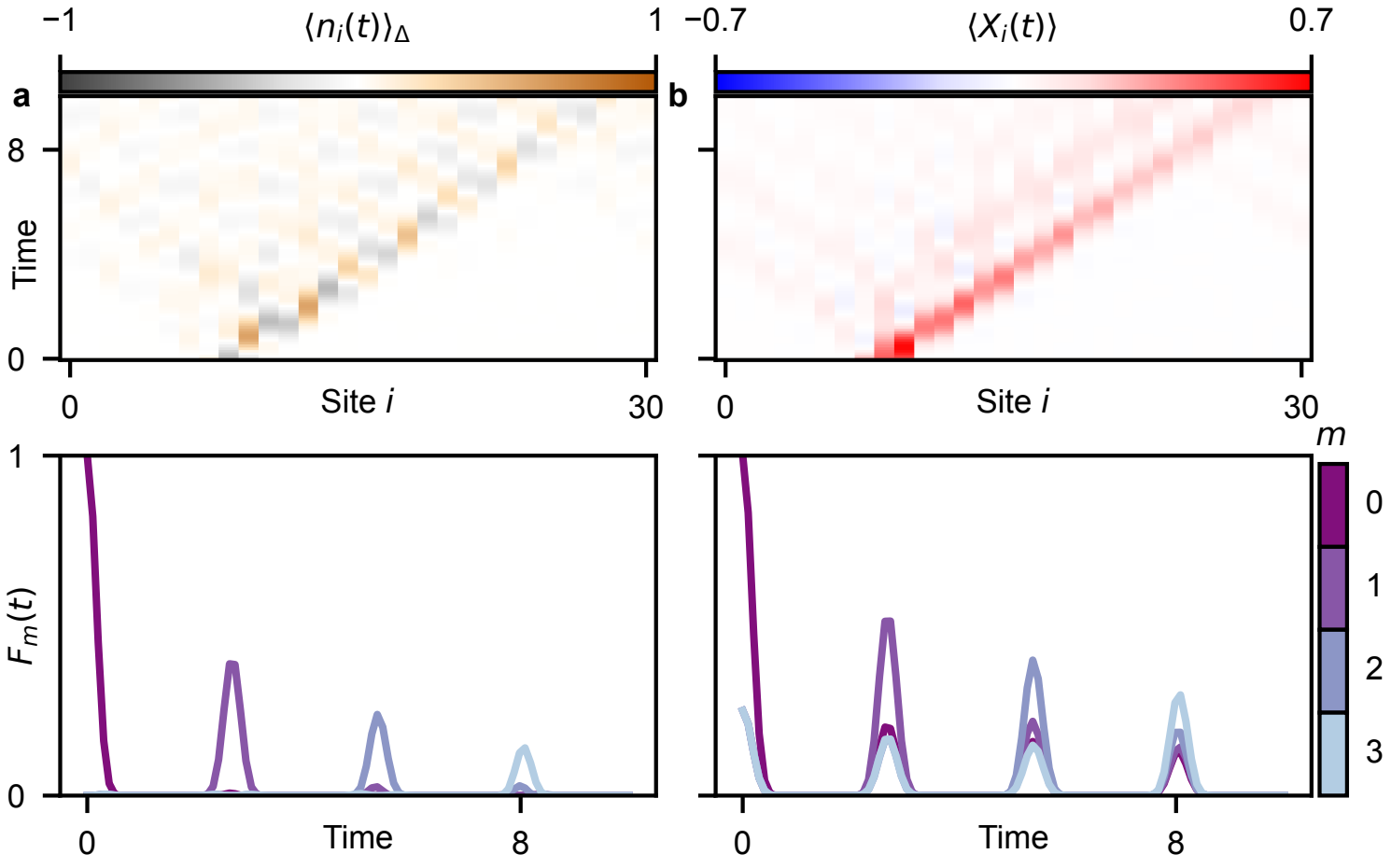

Supplementary Fig. 15. Unitary time-evolution of a system with 30 sites with periodic boundary conditions under the PPXPP Hamiltonian produced using Krylov time evolution. The states have a scarred background with the 5-site unit cell $|S_5\rangle$ and the defect cell for both states is positioned at the sites $\{6, 7, 8, 9, 10\}$. **a,** The zero-energy defect cell $|R_5\rangle$ initially shows coherent propagation to the right but distorts the scarred background more and also decays faster compared to the solitons in the PXP model. Also, the translation fidelity peaks show less pronounced revivals and decay faster **b,** Similar behavior can be observed for the energy-carrying soliton $|R_5^+\rangle$.

Hence, we search for the soliton cells by imposing similar constraints on the potential left- and right-moving defect cells as in the PXP model: (i) the first two sites have to be in the ground state to avoid blockade violation, (ii) $\{|S_5\rangle, |R_5\rangle, |L_5\rangle\}$ are orthonormal to each other and (iii) they satisfy the conjugation condition under chiral symmetry, $|L_5\rangle = \mathcal{C}\,|R_5\rangle$. We optimize over the family of states that satisfy these conditions to maximize the translation fidelity of a single right-moving defect $|R_5\rangle$ after one revival period. Numerical optimization yields the following cells:

$$|L_5, R_5\rangle = \frac{1}{c}|{\downarrow\downarrow}\rangle(\mp 1.133|{\downarrow\downarrow\downarrow}\rangle + 0.84i|{\uparrow\downarrow\downarrow}\rangle + 0.10i|{\downarrow\uparrow\downarrow}\rangle - 0.646i|{\downarrow\downarrow\uparrow}\rangle), \tag{28}$$

with $c$ being a normalization constant. These cells, taken in superposition with $|S_5\rangle$ in the form $|R_5^+\rangle = (|S_5\rangle + |R_5\rangle)/\sqrt{2}$, can also be used to initialize right- (and analogously left-) moving energy solitons.

We numerically check the unitary dynamics of such optimized cells now in a larger system. Specifically, inserting $|R_5\rangle$ in the scarred state $|K_5\rangle$ yields a right-moving defect as can be seen in Supplementary Fig. 15**a**. There, we show, similarly to the main text, the difference in expectation value of the local number operator between dynamics with and without a soliton defect. It is visible that although the propagation of disturbance is highly asymmetric between left and right directions, the pattern becomes washed out faster. Accordingly, the translation fidelity peaks are less pronounced compared to the PXP model. Finally, in Supplementary Fig. 15**b** we also see that the energy soliton $|R^+\rangle$ defined above shows somewhat more stable coherent propagation.

In conclusion, the results shown in this section suggest that solitons are not limited to the PXP model. This opens an exciting direction for searching for coherently propagating excitations in other models, with PPPXPPP and longer-range blockade generalizations of PXP being one perspective family.

## SUPPLEMENTARY REFERENCES

[1] Turner, C. J., Michailidis, A. A., Abanin, D. A., Serbyn, M. & Papić, Z. Weak ergodicity breaking from quantum many-body scars. *Nature Physics* **14**, 745–749 (2018).

[2] Kerschbaumer, A., Ljubotina, M., Serbyn, M. & Desaules, J.-Y. Quantum many-body scars beyond the PXP model in Rydberg simulators. *Physical Review Letters* **134**, 160401 (2025).

[3] Serbyn, M., Abanin, D. A. & Papić, Z. Quantum many-body scars and weak breaking of ergodicity. *Nature Physics* **17**, 675–685 (2021).

[4] Chandran, A., Iadecola, T., Khemani, V. & Moessner, R. Quantum many-body scars: A quasiparticle perspective. *Annual Review of Condensed Matter Physics* **14**, 443–469 (2023).

[5] Moudgalya, S., Bernevig, B. A. & Regnault, N. Quantum many-body scars and Hilbert space fragmentation: A review of exact results. *Reports on Progress in Physics* **85**, 086501 (2022).

[6] Virtanen, P. *et al.* SciPy 1.0: Fundamental Algorithms for Scientific Computing in Python. *Nature Methods* **17**, 261–272 (2020).

[7] Storn, R. & Price, K. Differential Evolution – A Simple and Efficient Heuristic for global Optimization over Continuous Spaces. *Journal of Global Optimization* **11**, 341–359 (1997). Publisher: Springer Science and Business Media LLC.

[8] Michailidis, A. A., Turner, C. J., Papić, Z., Abanin, D. A. & Serbyn, M. Slow quantum thermalization and many-body revivals from mixed phase space. *Physical Review X* **10**, 011055 (2020).

[9] Hu, Z. & Wu, B. Variational method for $\mathbb{Z}_K$ wavefunctions in spin-$J$ PXP model. *arXiv ePrints* (2025). 2501.09301.

[10] Wilkinson, J. W. P., Klobas, K., Prosen, T. & Garrahan, J. P. Exact solution of the Floquet-PXP cellular automaton. *Physical Review E* **102**, 062107 (2020).

[11] Iadecola, T. & Vijay, S. Nonergodic quantum dynamics from deformations of classical cellular automata. *Physical Review B* **102**, 180302 (2020).

[12] Giudici, G., Surace, F. M. & Pichler, H. Unraveling PXP many-body scars through Floquet dynamics. *Physical Review Letters* **133**, 190404 (2024).

[13] Bluvstein, D. *et al.* Controlling quantum many-body dynamics in driven Rydberg atom arrays. *Science* **371**, 1355–1359 (2021).